\documentclass[showpacs,preprint,superscriptaddress,amsmath,amssymb,noshowkeys]{revtex4-1}
\usepackage{graphicx}
\usepackage{dcolumn}
\usepackage{bm}
\usepackage{epsfig}
\usepackage{subfigure}
\usepackage{graphics}
\usepackage{amssymb}
\usepackage{amsthm}
\usepackage{amsmath}
\usepackage{array}
\usepackage{tabularx}
\usepackage{multirow}
\usepackage{amsmath}

\usepackage{caption}
\usepackage{algorithm}
\usepackage{algpseudocode}

\begin{document}

\title{Discrete unified gas kinetic scheme with force term for incompressible fluid flows}

\author{Chen Wu}
\affiliation{State Key Laboratory of Coal Combustion, Huazhong
University of Science and Technology, Wuhan 430074, China}

\author{Baochang Shi}
\email[Corresponding author]{(shibc@hust.edu.cn)}
\affiliation{School of Mathematics and Statistics, Huazhong
University of Science and Technology, Wuhan 430074, China}%

\author{Zhenhua Chai}
\affiliation{School of Mathematics and Statistics, Huazhong
University of Science and Technology, Wuhan 430074, China}%

\author{Peng Wang}
\affiliation{State Key Laboratory of Coal Combustion, Huazhong
University of Science and Technology, Wuhan 430074, China}

\date{\today}

\begin{abstract}
The discrete unified gas kinetic scheme (DUGKS) is a finite-volume scheme with discretization of particle velocity space, which combines the advantages of both lattice Boltzmann equation (LBE) method and unified gas kinetic scheme (UGKS) method, such as the simplified flux evaluation scheme, flexible mesh adaption and the asymptotic preserving properties. However, DUGKS is proposed for near incompressible fluid flows, the existing compressible effect may cause some serious errors in simulating incompressible problems. To diminish the compressible effect, in this paper a novel DUGKS model with external force is developed for incompressible fluid flows by modifying the approximation of Maxwellian distribution. Meanwhile, due to the pressure boundary scheme, which is wildly used in many applications, has not been constructed for DUGKS, the non-equilibrium extrapolation (NEQ) scheme for both velocity and pressure boundary conditions is introduced. To illustrate the potential of the proposed model, numerical simulations of steady and unsteady flows are performed. The results indicate that the proposed model can reduce the compressible effect efficiently against the original DUGKS model, and the NEQ scheme fits well with our model as they are both of second-order accuracy. We also implement the proposed model in simulating the three dimensional problem: cubical lid-driven flow. The comparisons of numerical solutions and benchmarks are presented in terms of data and topology. And the motion pattern of the fluid particles in a specific area is characterized for the steady-state cubical lid-driven flows.

\end{abstract}


\maketitle

\section{\label{sec:level1}Introduction}
\label{sec:introduction}
During the past decades, numerical methods based on kinetic theory have emerged as effective tools for computational fluid flows (CFD), such as the lattice Boltzmann equation (LBE) method \cite{succi2001lattice} and the unified gas kinetic scheme (UGKS) method \cite{xu2001gas, xu2010unified}. Different from the classical CFD methods, the LBE and UGKS methods simulate the fluid flows intrinsically in the mesoscopic scale, and this feature leads to many unique advantages in multi-scale and complex boundary problems \cite{guo2011velocity, guo2008lattice, zhang2006capturing, guo2009lattice, meng2011multiscale, guo2006physical}. Recently, based on the Boltzmann equation, the discrete unified gas-kinetic scheme (DUGKS) method has been proposed for isothermal flow by Guo et al. \cite{guo2013discrete} which combines the advantages of both LBE and UGKS methods: firstly, DUGKS has the simplified flux evaluation scheme and the conservative discrete collision operator, as the LBE; secondly, DUGKS is a finite-volume method with the adaption of flexible mesh and possessing the asymptotic preserving properties, as the UGKS. More details of DUGKS can be found in Ref. \cite{guo2013discrete}. 

Modeling for incompressible fluid flow has wide applications. However, the original DUGKS model is established for near incompressible flows, and it may cause some serious errors in simulating incompressible flows due to the compressible effect existing in the model. In this paper, we proposed a novel DUGKS model for incompressible fluid flows. To diminish the compressible effect, we modified the approximation of Maxwellian distribution which is inspired by the incompressible LBE model proposed by He and Luo \cite{he1997lattice}. Meanwhile, an external force, which is wildly used in many numerical simulations, is incorporated in the DUGKS model and also applies for the incompressible DUGKS model. Details will be shown in Section \ref{sec:numerical_methods}. 

Besides, the boundary conditions play important roles in numerical methods. In DUGKS method, the commonly used boundary conditions are the bounce-back (BB) scheme and the diffuse-scattering scheme. And these two schemes are applied for velocity boundary condition \cite{guo2013discrete}. However, the scheme for pressure boundary condition has not been constructed yet. Thus, the non-equilibrium extrapolation (NEQ) method \cite{zhao2002non} for both velocity and pressure boundary conditions is introduced into DUGKS.

To show the potential of the proposed model and the NEQ boundary scheme, comparisons between numerical solutions and analytical (or benchmark) results of both steady and unsteady flows are provided. In Section \ref{sec:Periodic_flow} and \ref{sec:Poiseuille}, the force driven periodic flow \cite{zou1995improved} and Poiseuille flow are simulated to illustrate the accuracy of the proposed model and the boundary schemes. In Section \ref{sec:Womersley}, compared with the original DUGKS model, the simulation of unsteady Womersley flow shows the compressible effect reduction of the proposed model. And in Section \ref{sec:2D_LDF}, the 2D-LDF at different Reynold numbers $Re=400,1000,5000,7500$ are simulated to show the application of the non-uniform mesh. It also shows the accuracy and robustness of the proposed model when the Reynold number reaches $Re=12000$.

Finally, the proposed model is implemented to simulate the three dimensional lid-driven flow. The comparisons of data and topology between numerical results and benchmarks are presented in Section \ref{sec:3D_LDF}. And in steady state cubical lid-driven flow, the motion pattern of the fluid particles near the focus point \cite{sheu2002flow} is found by analyzing the numerical solutions topologically.

\section{Numerical methods} 
\label{sec:numerical_methods}

\subsection{DUGKS for incompressible fluid flows}
The DUGKS model is starting from the Boltzmann equation with Bhatnagar-Gross-Krook (BGK) collision model \cite{bhatnagar1954model},
\begin{equation}
\frac{{\partial f}}{{\partial t}} + \bm{\xi}  \cdot \nabla f = \Omega,
\end{equation}
where $f = f({\bm{x}},\bm{\xi} ,t)$ is the particle distribution function for particles at position $\bm{x}$ and time $t$ moving with velocity $\bm{\xi}$, $\tau$ is the relaxation time, and $\Omega$ is the collision term, which is given by:
\begin{equation}{\label{collision term}}
\Omega ({\bm{x}},\bm{\xi} ,t) =  - \frac{1}{\tau }\left[ {f({\bm{x}},\bm{\xi} ,t) - {f^{eq}}({\bm{x}},\bm{\xi} ,t)} \right].
\end{equation}
In addition, $f^{eq}$ is the Maxwellian equilibrium distribution function,
\begin{equation}
{f^{eq}} = \frac{\rho }{{{{(2\pi RT)}^{D/2}}}}\exp \left(  - \frac{{{\left| {\bm{\xi}  - \bm{u}} \right|}^2}}{{2RT}} \right),
\end{equation}
in which $\rho$ is the density, $R$ is the gas constant, $T$ is the temperature, $D$ is the spatial dimension, and $\bm{u}$ is the fluid velocity.

As the DUGKS model is a finite-volume method, we divide the flow domain into a set of control volumes, and each control volume $V_j$ is centered at $\bm{x}_j$. Then, its evolution equation can be obtained as:
\begin{equation}
{f_i}^{n + 1} - {f_i}^n + \frac{{\Delta t}}{{\left| {{V_j}} \right|}}{F^{n + 1/2}} = \frac{{\Delta t}}{2}[{\Omega _j}^{n + 1} + {\Omega _j}^n],
\end{equation}
where $\Delta t$ is time step, $f^{n}_j$ and $\Omega^{n}_j$ are the cell-averaged values of the distribution function and collision term, and $F^{n}$ is the microflux across the cell interface. Details about this evolution equation will be discussed in Section \ref{sec:force_term}. 

In this section, we focus on the Maxwellian equilibrium distribution function $f^{eq}$. For the DUGKS model, $f^{eq}$ is approximated by its second-order Taylor or Hermit expansion of the Mach number $Ma \approx \left| \textbf{u} \right|/\sqrt {RT}  \ll 1.0$. And then, by discretizing the velocity space and employing the Gauss-Hermit quadrature, the equilibrium distribution is modified into discrete form \cite{he1997priori, luo2000some, shan2006kinetic} as follows:
\begin{equation}{\label{feq}}
f_i^{eq} = {W_i}\rho \left[ {1 + \frac{{\bm{\xi}_i  \cdot \textbf{u}}}{{RT}} + \frac{{{{(\bm{\xi}_i  \cdot \textbf{u})}^2}}}{{2{{(RT)}^2}}} - \frac{{{{\left| \textbf{u} \right|}^2}}}{{2RT}}} \right]
\end{equation}
where $W_i$ is the weights determined by the abscissas $\bm{\xi}_i$, as shown in Sec. \ref{sec:discrete_velocities}.

However, in the case of incompressible flow, the compressible effect in the original DUGKS model should be noticed which may cause some significant errors in numerical simulations. Thus, aiming at reducing or eliminating the compressible effect, we proposed the incompressible DUGKS model for both steady and unsteady flows.

For incompressible flow, the density can be seen as $\rho = \rho_0 + \Delta \rho$, where $\rho_0$ is the approximate constant density of fluid, and $\Delta \rho$ is the density fluctuation which should be of the order $O(Ma^{2})$\cite{he1997lattice, zou1995improved}.
Thus, we can introduce a new type of equilibrium distribution function for DUGKS,
\begin{equation}{\label{icfeq}}
f_i^{eq} = {W_i}\left\{ {\rho  + {\rho _0}\left[ {\frac{{\bm{\xi}_i  \cdot \textbf{u}}}{{RT}} + \frac{{{{(\bm{\xi}_i  \cdot \textbf{u})}^2}}}{{2{{(RT)}^2}}} - \frac{{{{\left| \textbf{u} \right|}^2}}}{{2RT}}} \right]} \right\},
\end{equation}
which is inspired by the incompressible lattice Boltzmann model\cite{he1997lattice}. It can be easily found that the neglected terms such as $\Delta \rho (\bm{u}/\sqrt{RT})$ are of the order $O(Ma^3)$ or higher. With this equilibrium distribution function, the fluid density $\rho$ and the velocity $\bm{u}$ can be obtained by

\begin{subequations}
\begin{equation}
\rho  = \sum\limits_i {{f_i}},
\end{equation}
\begin{equation}
{\rho _0}\bm{u} = \sum\limits_i {{\bm{\xi} _i}{f_i}}.
\end{equation}
\end{subequations}

\subsection{Incompressible DUGKS model with force term}
\label{sec:force_term}
In this section, an external force will be introduced in the DUGKS model and then applied for the proposed incompressible DUGKS model.

At the beginning, we consider the Boltzmann equation with a force term $S$ \cite{he1998novel},
\begin{equation}{\label{BoltzmannEquationWithForceTerm}}
{\partial _t}f + \bm{\xi}  \cdot \nabla f =  \Omega + S,
\end{equation}
where
\begin{equation}
\label{S}
S = \frac{{\bm{G} \cdot (\bm{\xi}  - \bm{u)}}}{{RT}}{f^{eq}},
\end{equation}
with $\bm{G}$ being the acceleration, and $f^{eq}$ is the Maxwellian equilibrium distribution function. By integrating Eq. \eqref{BoltzmannEquationWithForceTerm} on $V_j$ from time $t_n$ to $t_{n+1} = t_n + \Delta t$, and using the midpoint rule for the integration of the convection term and the trapezoidal rule for the collision term and the force term, it can be derived as
\begin{equation}{\label{eqOriginalEvolution}}
f_j^{n + 1} - f_j^n + \frac{{\Delta t}}{{\left| {{V_j}} \right|}}{F^{n + 1/2}} = \frac{{\Delta t}}{2}\left[ {\Omega _j^{n + 1} + \Omega _j^n} \right] + \frac{{\Delta t}}{2}\left[ {S_j^{n + 1} + S_j^n} \right],
\end{equation}
where the cell-averaged values of the distribution function $f_j^n$ and force term $S_j^n$ at the time $t_n$ are given by
\begin{equation}
f_j^n = \frac{1}{{\left| {{V_j}} \right|}}\int_{{V_j}} {f(\bm{x},\bm{\xi} ,{t_n})} {\kern 1pt} d\bm{x},
\end{equation}
\begin{equation}
S_j^n = \frac{1}{{\left| {{V_j}} \right|}}\int_{{V_j}} {S(\bm{x},\bm{\xi} ,{t_n})} {\kern 1pt} d\bm{x},
\end{equation}
at the same time, the microflux across the cell interface is given by
\begin{equation}
\label{microflux}
{F^{n+1/2}} = \int_{\partial {V_j}} {(\bm{\xi}  \cdot \bm{n})f(\bm{x},\bm{\xi} ,{t_n})} {\kern 1pt} d\bm{x},
\end{equation}
in which ${\left| {{V_j}} \right|}$ and ${\partial {V_j}}$ are the volume and surface of cell $V_j$, and $\bm{n}$ is the outward unit vector normal to the surface.

Clearly, the form of Eq. \eqref{eqOriginalEvolution} is implicit. For the purpose of gaining its explicit form,
we define a new distribution function,
\begin{equation}
\tilde f = f - \frac{{\Delta t}}{2}\Omega  - \frac{{\Delta t}}{2}S,
\end{equation}
So, Eq. \eqref{eqOriginalEvolution} can be rewritten explicitly as
\begin{equation}
\tilde f_j^{n + 1} = \left( {1 - \frac{{2\Delta t}}{{2\tau  + \Delta t}}} \right)\tilde f_j^n + \frac{{2\Delta t}}{{2\tau  + \Delta t}}{f^{eq}} + \frac{{2\tau \Delta t}}{{2\tau  + \Delta t}}S - \frac{{\Delta t}}{{\left| {{V_j}} \right|}}{F^{n + 1/2}}.
\end{equation}
Furthermore, we can also obtain the same equation of the DUGKS in form,
\begin{equation}
\label{evolution_equation}
\tilde f_j^{n + 1} = \tilde f_j^{ + ,n} - \frac{{\Delta t}}{{\left| {{V_j}} \right|}}{F^{n + 1/2}},
\end{equation}
only if the term $\tilde f_j^{ + ,n}$ is defined as
\begin{equation}
\label{f_j+n}
{\tilde f^ + } = \frac{{2\tau  - \Delta t}}{{2\tau  + \Delta t}}\tilde f + \frac{{2\Delta t}}{{2\tau  + \Delta t}}{f^{eq}} + \frac{{2\tau \Delta t}}{{2\tau  + \Delta t}}S
\end{equation}

Then, the distribution function at the cell interface is also needed for evaluating the flux $F^{n+1/2}$. So we integrate the Boltzmann equation along the characteristic line to the end point $x_b$ located at the cell interface within a half time step $h=\Delta t / 2$,
\begin{equation}
\begin{array}{l}
f({\bm{x_b}},\bm{\xi} ,{t_n} + h) - f({\bm{x_b}} - \bm{\xi} h,\bm{\xi} ,{t_n})\\
 = \frac{h}{2}\left[ {\Omega ({\bm{x_b}},\bm{\xi} ,{t_n} + h) + \Omega ({\bm{x_b}} - \bm{\xi} h,\bm{\xi} ,{t_n})} \right] + \frac{h}{2}\left[ {S({\bm{x_b}},\bm{\xi} ,{t_n} + h) + S({\bm{x_b}} - \bm{\xi} h,\bm{\xi} ,{t_n})} \right],
\end{array}
\end{equation}
and use trapezoidal rule again to treat the collision and force terms,
\begin{equation}{\label{DF_INTER}}
\bar f({\bm{x}_b},\bm{\xi} ,{t_n} + h) = \frac{{2\tau  - h}}{{2\tau  + h}}\bar f({\bm{x}_b} - \bm{\xi} h,\bm{\xi} ,{t_n}) + \frac{{2h}}{{2\tau  + h}}{f^{eq}} + \frac{{2\tau h}}{{2\tau  + h}}S,
\end{equation}
where
\begin{equation}
\label{original_f}
\bar f = f - \frac{{h}}{2}\Omega  - \frac{{h}}{2}S = \frac{{2\tau  + h}}{{2\tau }}f - \frac{h}{{2\tau }}{f^{eq}} - \frac{h}{2}S
\end{equation}
Also, the Eq. \eqref{DF_INTER} can be rewritten as
\begin{equation}{\label{interfaceEvof}}
\bar f({\bm{x}_b},\bm{\xi} ,{t_n} + h) = {\bar f^ + }({\bm{x}_b} - \bm{\xi} h,\bm{\xi} ,{t_n}),
\end{equation}
where
\begin{equation}
{\bar f^ + } = \frac{{2\tau  - h}}{{2\tau  + h}}\bar f + \frac{{2h}}{{2\tau  + h}}{f^{eq}} + \frac{{2\tau h}}{{2\tau  + h}}S.
\end{equation}
It can be observed that Eq. \eqref{interfaceEvof} is similar to the corresponding equation in the original DUGKS model. Consequently, we can get $\bar f$ in the same way,
\begin{equation}
\label{gradient}
\bar f({\bm{x}_b},\bm{\xi} ,{t_n} + h) = {\bar f^ + }({\bm{x}_b},\bm{\xi} ,{t_n}) - \bm{\xi} h \cdot \bm{\sigma _b},
\end{equation}
where $\bm{\sigma _b} = \nabla {\bar f^ + }({x_b},\xi ,{t_n})$ is the gradient. Additionally, the relationships among ${\tilde f}$, ${\tilde f^ + }$ and ${\bar f^ + }$ will be used in computation:
\begin{equation}
\label{f_bar_+}
{\bar f^ + } = \frac{{2\tau  - h}}{{2\tau  + \Delta t}}\tilde f + \frac{{3h}}{{2\tau  + \Delta t}}{f^{eq}} + \frac{{3\tau h}}{{2\tau  + \Delta t}}S,
\end{equation}
\begin{equation}
\label{f_tilde_+}
{\tilde f^ + } = \frac{{4}}{{3}}\bar f^+ - \frac{{1}}{{3}} \tilde f.
\end{equation}

After discretizing particle velocity space and employing the corresponding equilibrium distribution function Eq. \eqref{feq}, the fluid density $\rho$ and the velocity $\bm{u}$ should be computed discretely from
\begin{subequations}
\begin{equation}
\rho  = \int {\tilde f{\kern 1pt} d} \bm{\xi},
\end{equation}
\begin{equation}
\rho \bm{u} = \int {\bm{\xi} \tilde f{\kern 1pt} d\bm{\xi} }  + \frac{{\rho G\Delta t}}{2}.
\end{equation}
\end{subequations}

To summarize, the DUGKS model with force term is established, and one can update $\tilde f$ from $t$ to $t + \Delta t$ with the algorithm presented in Appendix \ref{sec:appendix}.

For the proposed incompressible DUGKS model, all the above processes are applicable, but there are differences should be claimed. Since we modified the equilibrium distribution function as Eq. \eqref{icfeq}, the conservative flow variables should be computed by
\begin{subequations}
\label{macro_values}
\begin{equation}
\rho  = \int {\tilde f{\kern 1pt} d} \bm{\xi},
\end{equation}
\begin{equation}
\rho_0 \bm{u} = \int {\bm{\xi} \tilde f{\kern 1pt} d\bm{\xi} }  + \frac{{\rho_0 G\Delta t}}{2},
\end{equation}
\end{subequations}
where the terms of $\Delta \rho$ $(Ma^2)$ are ignored. As we mentioned above, under the incompressible limit, the accuracy of our model is still guaranteed.

\subsection{Discrete particle velocities}
\label{sec:discrete_velocities}
By using the three-point Gauss-Hermite quadrature \cite{guo2013discrete}, the discrete particle velocities and associated weights are given by,
\begin{equation}
{\bm{\xi}} = \sqrt {3RT} \left[ {\begin{array}{*{20}{c}}
{ - 1}& \ 0& \ 1
\end{array}} \right], \
{W_i} = \left\{ {\begin{array}{*{20}{l}}
{\frac{2}{3},}&{i = 2}\\
{\frac{1}{3},}&{i = 1,3}
\end{array}} \right. .
\end{equation}
Using the tensor product method, the discrete velocities and weights for higher-dimensional flows are generated as follows.

\begin{figure}
\includegraphics[width=2.7in,height=2.7in]{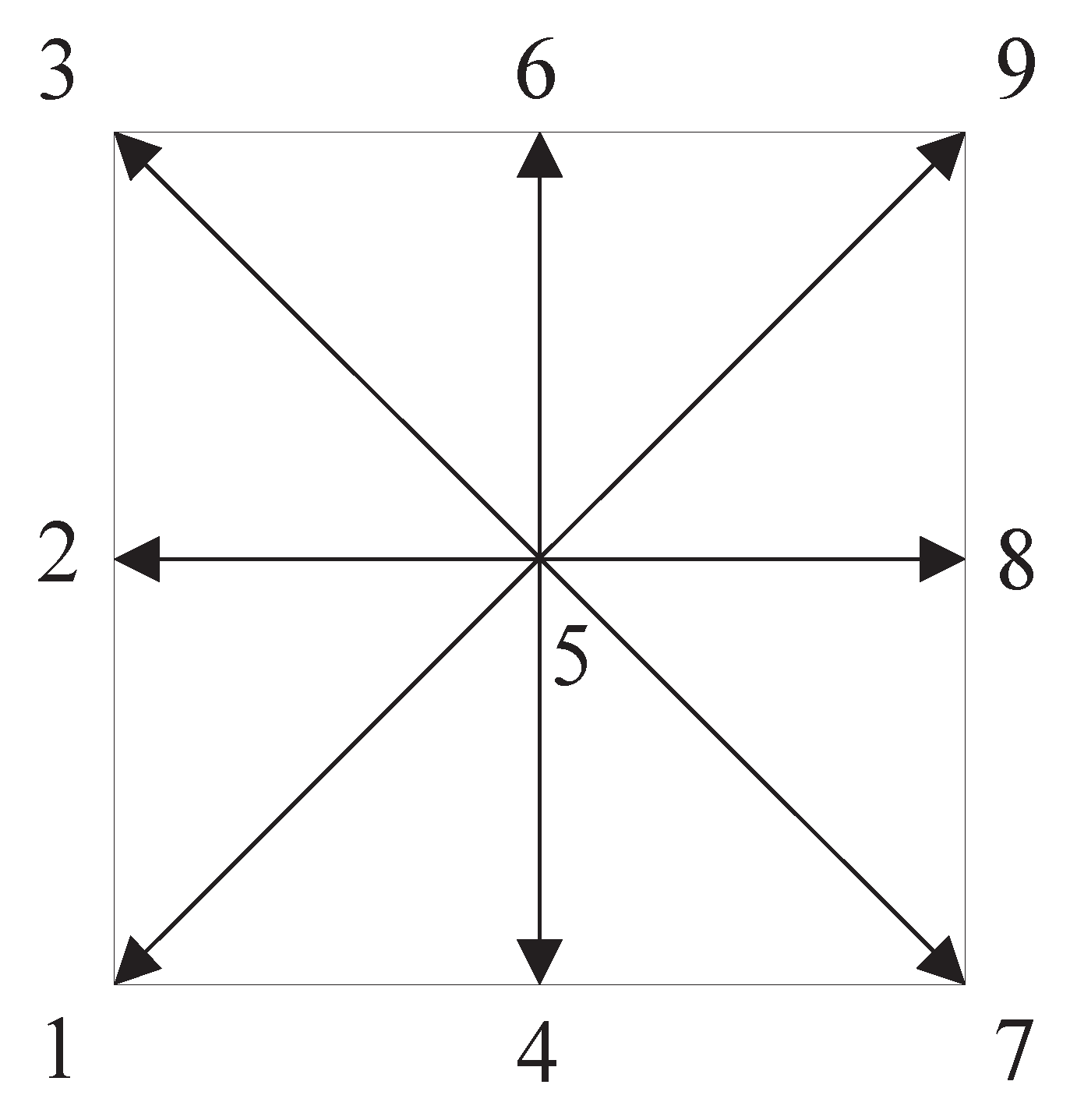}
\caption{The discrete velocities of two dimensional DUGKS model.} 
\label{fig:DiscreteVelocity_2D}
\end{figure}

For 2D flows, the discrete velocities and associated weights can be computed as,
\begin{equation}
{\bm{\xi}} = \sqrt {3RT} \left[ {\begin{array}{*{20}{c}}
{ - 1}&{ - 1}&{ - 1}& \ 0& \ 0& \ 0& \ 1& \ 1& \ 1\\
{ - 1}& \ 0& \ 1&{ - 1}& \ 0& \ 1&{ - 1}& \ 0& \ 1
\end{array}} \right], \
{W_i} = \left\{ {\begin{array}{*{20}{l}}
{\frac{4}{9},}&{i = 5}\\
{\frac{1}{9},}&{i = 2,4,6,8}\\
{\frac{1}{{36}},}&{i = 1,3,7,9}
\end{array}} \right. .
\end{equation}
The 9 discrete velocity vectors are shown in Fig. \ref{fig:DiscreteVelocity_2D}.

\begin{figure}
\includegraphics[width=2.7in,height=2.7in]{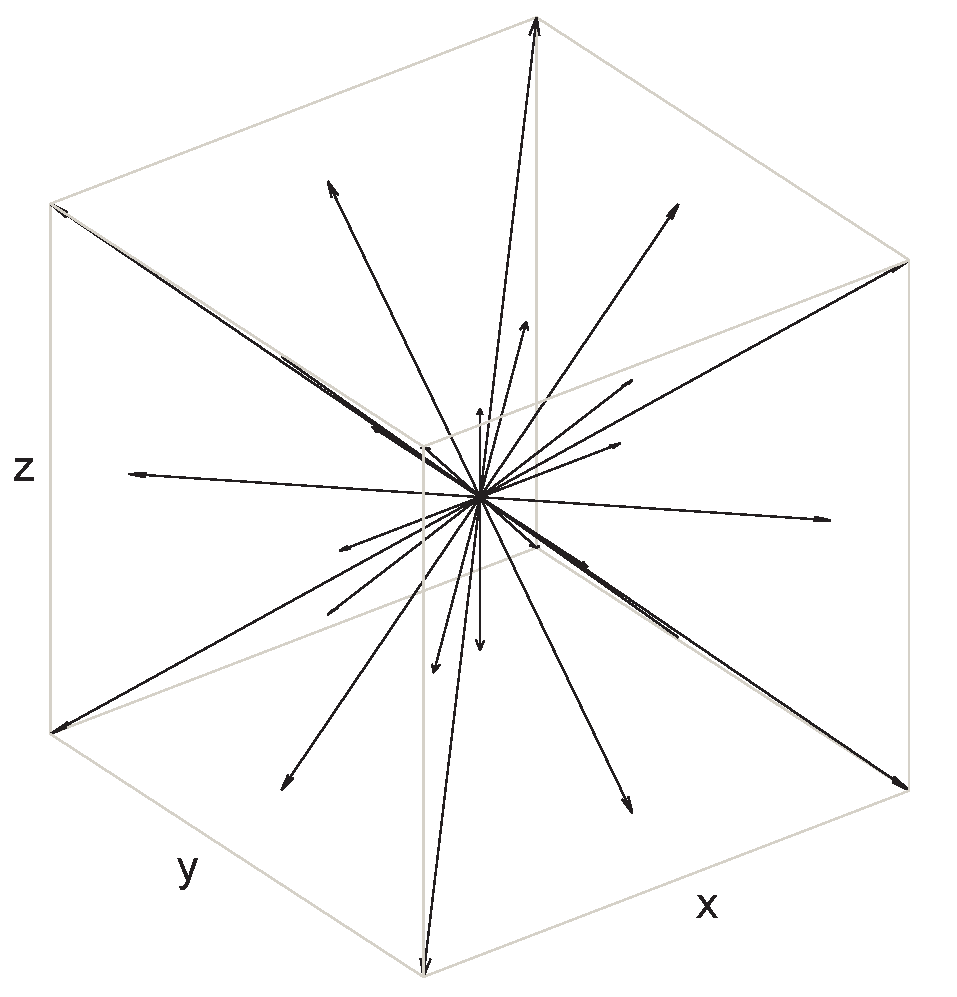}
\caption{The discrete velocities of three dimensional DUGKS model.} 
\label{fig:DiscreteVelocity_3D}
\end{figure}

Meanwhile, for 3D flows, we use the tensor product method again, and the discrete velocities and associated weights are given by,
\begin{equation}
\begin{split}
{\bm{\xi}} = \\
 c &\left[ \begin{smallmatrix}
{ - 1}&{ - 1}&{ - 1}& { - 1}&{ - 1}&{ - 1}& { - 1}&{ - 1}&{ - 1}& \ 0& \ 0& \ 0& \ 0& \ 0& \ 0& \ 0& \ 0& \ 0& \ 1& \ 1& \ 1& \ 1& \ 1& \ 1&\ 1& \ 1& \ 1\\
{ - 1}&{ - 1}&{ - 1}& \ 0& \ 0& \ 0& \ 1& \ 1& \ 1& \ { - 1}&{ - 1}&{ - 1}& \ 0& \ 0& \ 0& \ 1& \ 1& \ 1&\ { - 1}&{ - 1}&{ - 1}& \ 0& \ 0& \ 0& \ 1& \ 1& \ 1\\
{ - 1}& \ 0& \ 1&{ - 1}& \ 0& \ 1&{ - 1}& \ 0& \ 1& \ { - 1}& \ 0& \ 1&{ - 1}& \ 0& \ 1&{ - 1}& \ 0& \ 1& \ { - 1}& \ 0& \ 1&{ - 1}& \ 0& \ 1&{ - 1}& \ 0& \ 1
\end{smallmatrix} \right], \\
& c = \sqrt{3RT}, \ \ {W_i} = \left\{ {\begin{array}{*{20}{l}}
{\frac{8}{27},}&{i = 14}\\
{\frac{2}{27},}&{i = 5, 11, 13, 15, 17, 23}\\
{\frac{1}{{54}},}&{i = 2, 4, 6, 8, 10, 12, 16, 18, 20, 22, 24, 26}\\
{\frac{1}{{216}},}&{i = 1, 3, 7, 9, 19, 21, 25, 27}
\end{array}} \right. .
\end{split}
\end{equation}
The 27 discrete velocity vectors are illustrated in Fig. \ref{fig:DiscreteVelocity_3D}.

\subsection{Boundary conditions}
We will provide two types of boundary conditions in this section: the bounce-back (BB) method and the non-equilibrium extrapolation (NEQ) method \cite{zhao2002non}.

The bounce-back method is a commonly used boundary condition, and it assumes that velocity of the particle will just reverse when hitting the wall \cite{ladd1994numerical}. For those particles leaving the wall which is assumed locating at a cell interface $\bm{x}_w$, the distribution functions are given by
\begin{equation}
\label{BB}
\begin{split}
f({{\bm{x}}_w},{{\bm{\xi }}_i},t + h)& = f({{\bm{x}}_w}, - {{\bm{\xi }}_i},t + h) + 2{\rho _w}{W_i}\frac{{{{\bm{\xi }}_i} \cdot {{\bm{u}}_w}}}{{RT}},\\
&{{\bm{\xi }}_i} \cdot {\bm{n}} < 0,         
\end{split}
\end{equation}
where $\bm{n}$ is the outward unit vector normal to the wall, $\bm{u}_w$ is the wall velocity, and $\rho_w$ is the density of the wall which can be approximated by $\rho_0$ under the incompressible limit. 

Obviously, the bounce-back method is suitable for velocity boundary condition but not for pressure boundary condition which is also widely used. Thus, we introduce the non-equilibrium extrapolation method  \cite{zhao2002non} into the DUGKS model which can deal with both velocity and pressure boundary conditions.

\begin{figure}
\centering\includegraphics[width=2.0in]{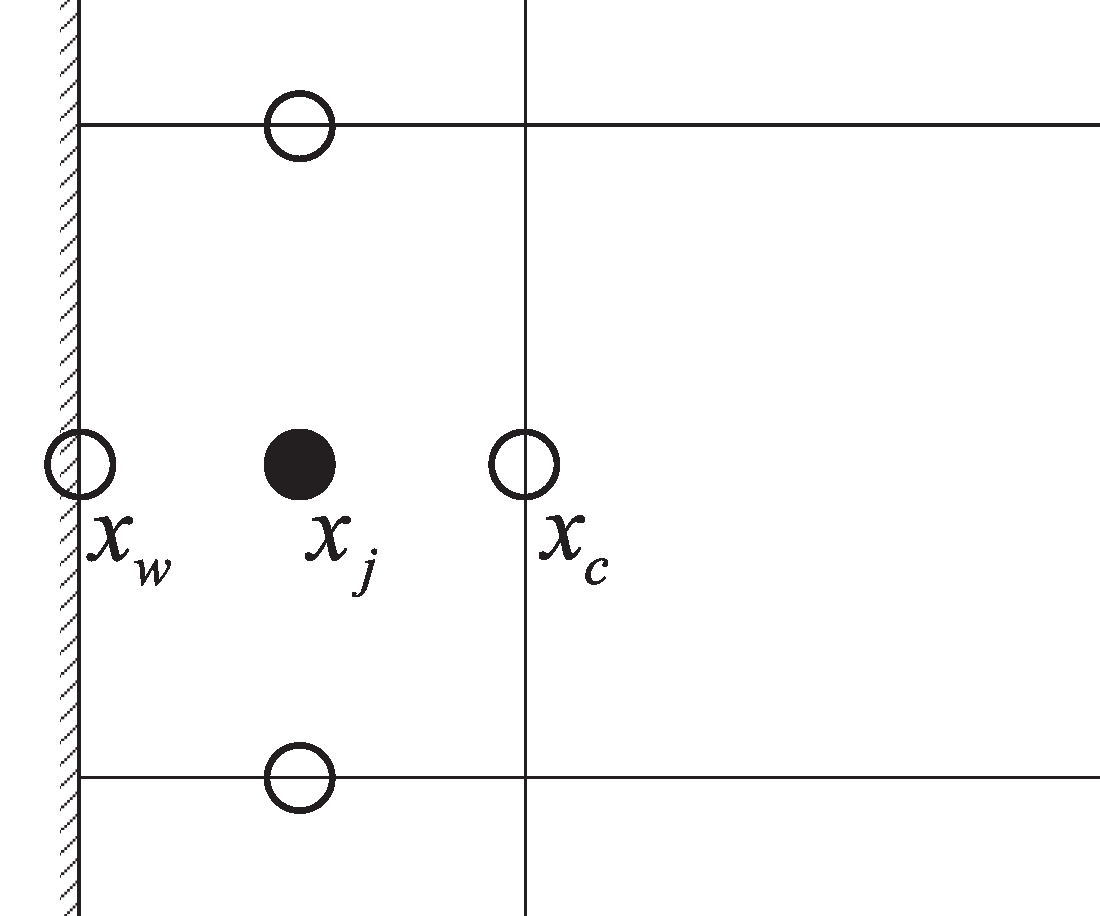} 
\caption{The point at cell center is ($\bullet$), and the point at cell interface is ($\circ$).}
\label{fig:Boundary} 
\end{figure}

The non-equilibrium extrapolation method can determine the distribution function at the wall from the given macroscopic values such as velocity or pressure. The distribution functions of the particle reflect from the wall are as follows,
\begin{equation}
\label{NEQ}
\begin{split}
f({\bm{x}_w},{\bm{\xi} _i},t + h) &= {f^{eq}}({\bm{\xi} _i}; \rho_{\alpha}, \bm{u}_{\alpha}) + {f^{neq}}({\bm{x}_w},{\bm{\xi} _i},t + h),\\
&{{\bm{\xi }}_i} \cdot {\bm{n}} < 0,  
\end{split}
\end{equation}
where the non-equilibrium part $f^{neq}$ can be approximated by the information of the point $\bm{x}_c$ next to $\bm{x}_w$ and at interface of the same cell shown in Fig. \ref{fig:Boundary},
\begin{equation}
{f^{neq}}({\bm{x}_w},{\bm{\xi} _i},t + h) = f({\bm{x}_c},{\bm{\xi} _i},t + h) - {f^{eq}}({\bm{\xi} _i};{\rho _c},{\bm{u}_c}).
\end{equation}
Besides, $\bm{u}_{\alpha}$ and $\rho_{\alpha}$ are velocity and density at the wall determined by different boundary conditions: for velocity condition, $\bm{u}_{\alpha} = \bm{u}_w$ and $\rho_{\alpha}$ is approximated by $\rho(\bm{x}_c, t+h)$; oppositely, for the pressure condition, $\rho_{\alpha}$ is computed from the pressure of the wall and $\bm{u}_{\alpha}$ is approximated by $\bm{u}(\bm{x}_c, t+h)$. The non-equilibrium extrapolation method which is also wildly used in LBE, is a second-order scheme, and this will be illustrated in next section.

\section{Numerical Results}
\label{sec:numerical_results}
In this section, five examples are simulated to show the capacity of the incompressible DUGKS model, where the periodic flow, Poiseuille flow and Womersley flow are of analytical solutions, and the lid-driven flows are of benchmarks. The contents are arranged as that mentioned in Section \ref{sec:introduction}.

In our simulations, the time step $\Delta t$ is determined by the Courant-Friedrichs-Lewy (CFL) condition which is given by
\begin{equation}
\Delta t = a \frac{{\Delta x}}{C},
\end{equation}
where $a$ is the CFL number, $\Delta x$ is the minimum grid spacing and $C$ is the maximum discrete velocity which is $\sqrt{3RT}$. Besides, the Mach number is computed by $Ma = U / C$ in our simulations.

\subsection{Periodic flow}
\label{sec:Periodic_flow}
The periodic flow, which is driven by extern force, is an incompressible and time-independent problem with an analytical solution \cite{zou1995improved}. We simulate this flow for testing the accuracy and convergence order of the incompressible DUGKS model. As its domain is periodic, the numerical results will not be influenced by boundary conditions. The analytical solution of the periodic flow is given by
\begin{subequations} \label{PF_AS}
\begin{equation}
u(x, y) = {u_0}\sin 2\pi x\sin 2\pi y,
\end{equation}
\begin{equation}
v(x, y) = {u_0}\cos 2\pi x\cos 2\pi y,
\end{equation}
\begin{equation}
p(x,y) = \frac{1}{4}{\rho _{0}}u_{0}^{2}(cos4\pi x - \cos 4\pi y),
\end{equation}
\end{subequations}
and the body force is given by
\begin{subequations}
\begin{equation}
{F_x}(x,y) = 8{\pi ^2}\nu {u_0}\sin 2\pi x\sin 2\pi y,
\end{equation}
\begin{equation}
{F_y}(x,y) = 8{\pi ^2}\nu {u_0}\cos 2\pi x\cos 2\pi y,
\end{equation}
\end{subequations}
where $ u_0 $, $\rho_0$ are constants, $\nu$ is the kinetic viscosity, $p$ is the pressure, $\bm{u} = (u,v)$ is the velocity, and the computation region is $x \in [0, 1]$, $y \in [0, 1]$.

\begin{figure}
\subfigure[]{
\label{fig:PF_velocity:0.3}
\includegraphics[width=2.7in,height=2.3in]{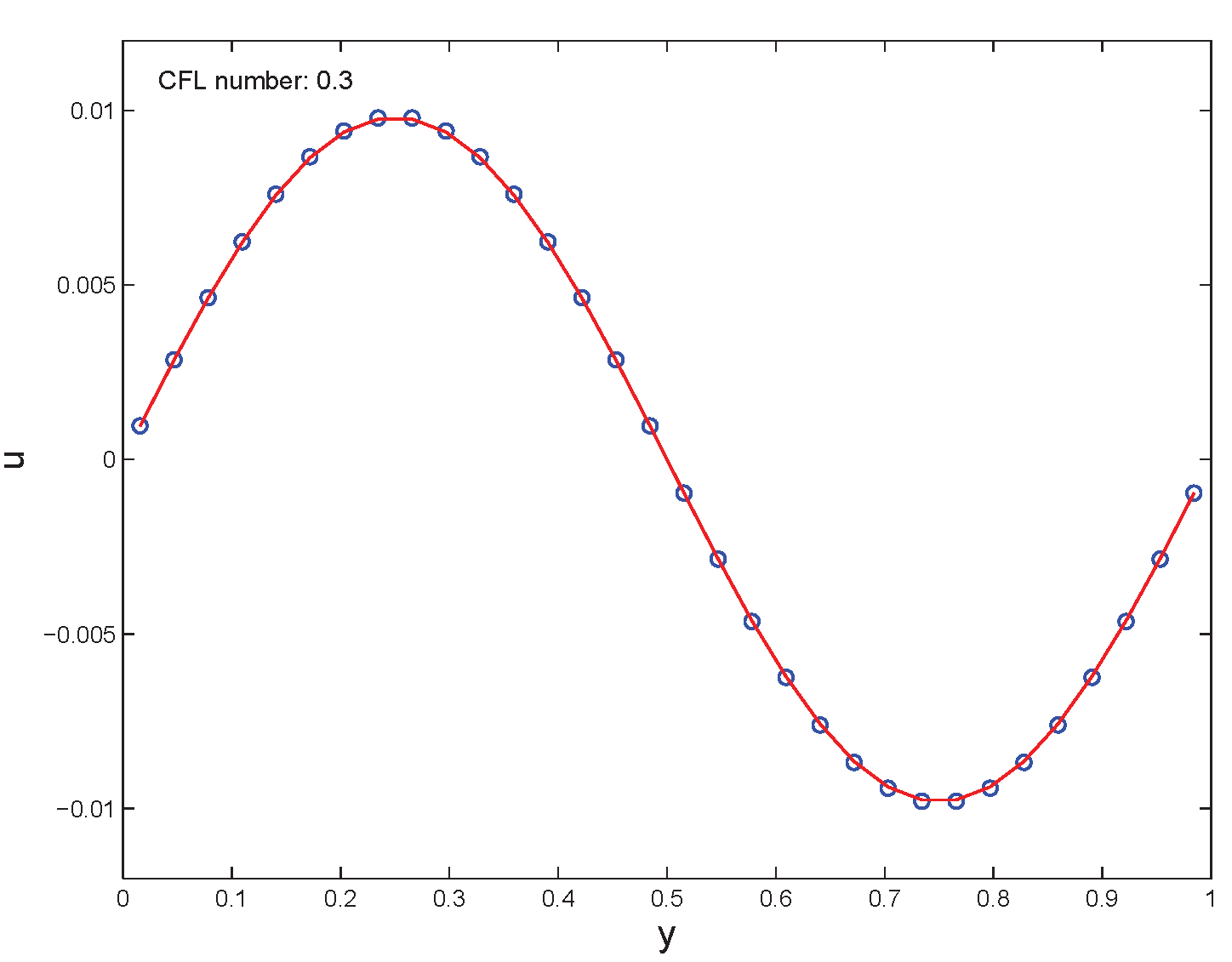}}
\subfigure[]{
\label{fig:PF_pressure:0.3}
\includegraphics[width=2.7in,height=2.3in]{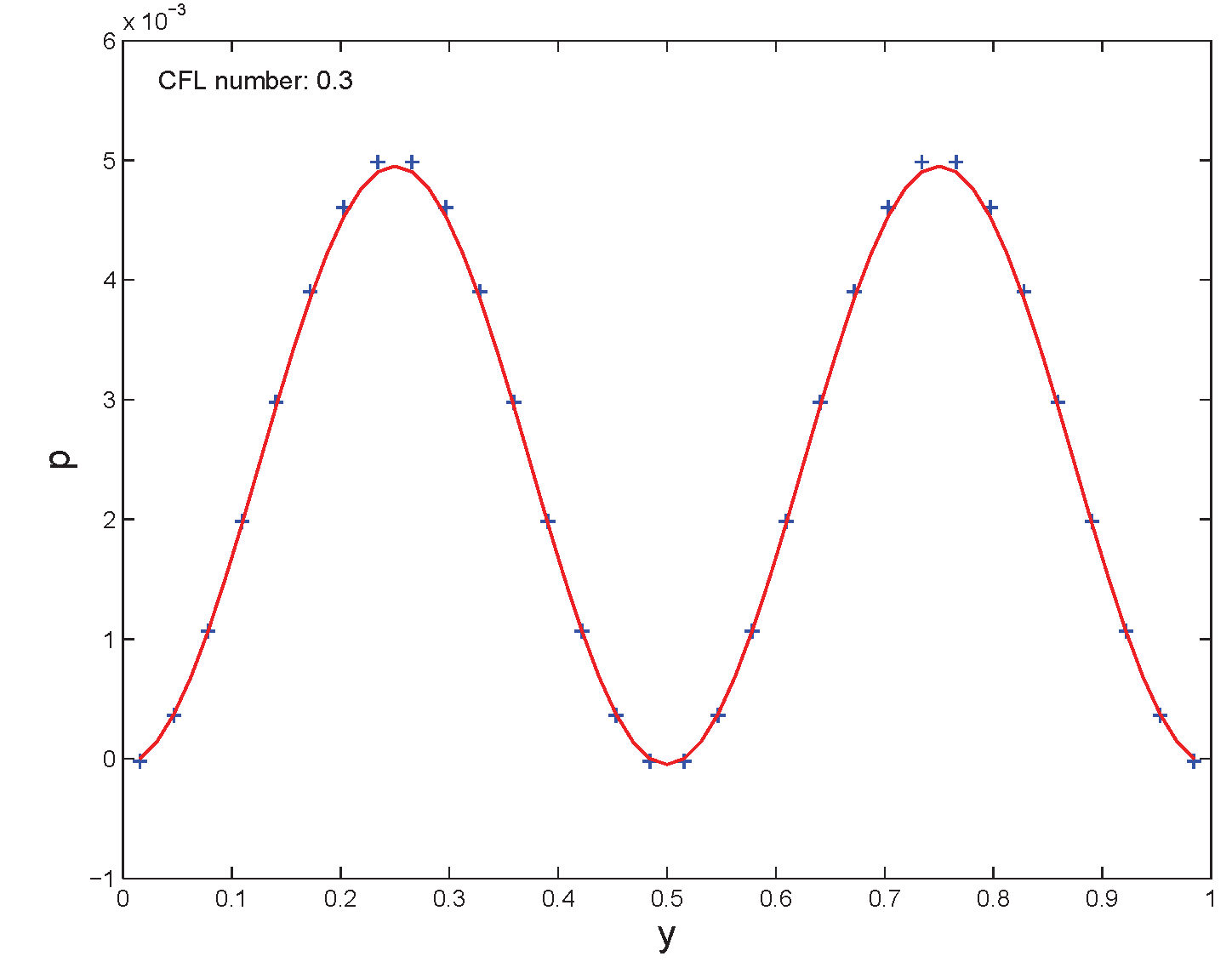}}
\subfigure[]{
\label{fig:PF_velocity:0.5}
\includegraphics[width=2.7in,height=2.3in]{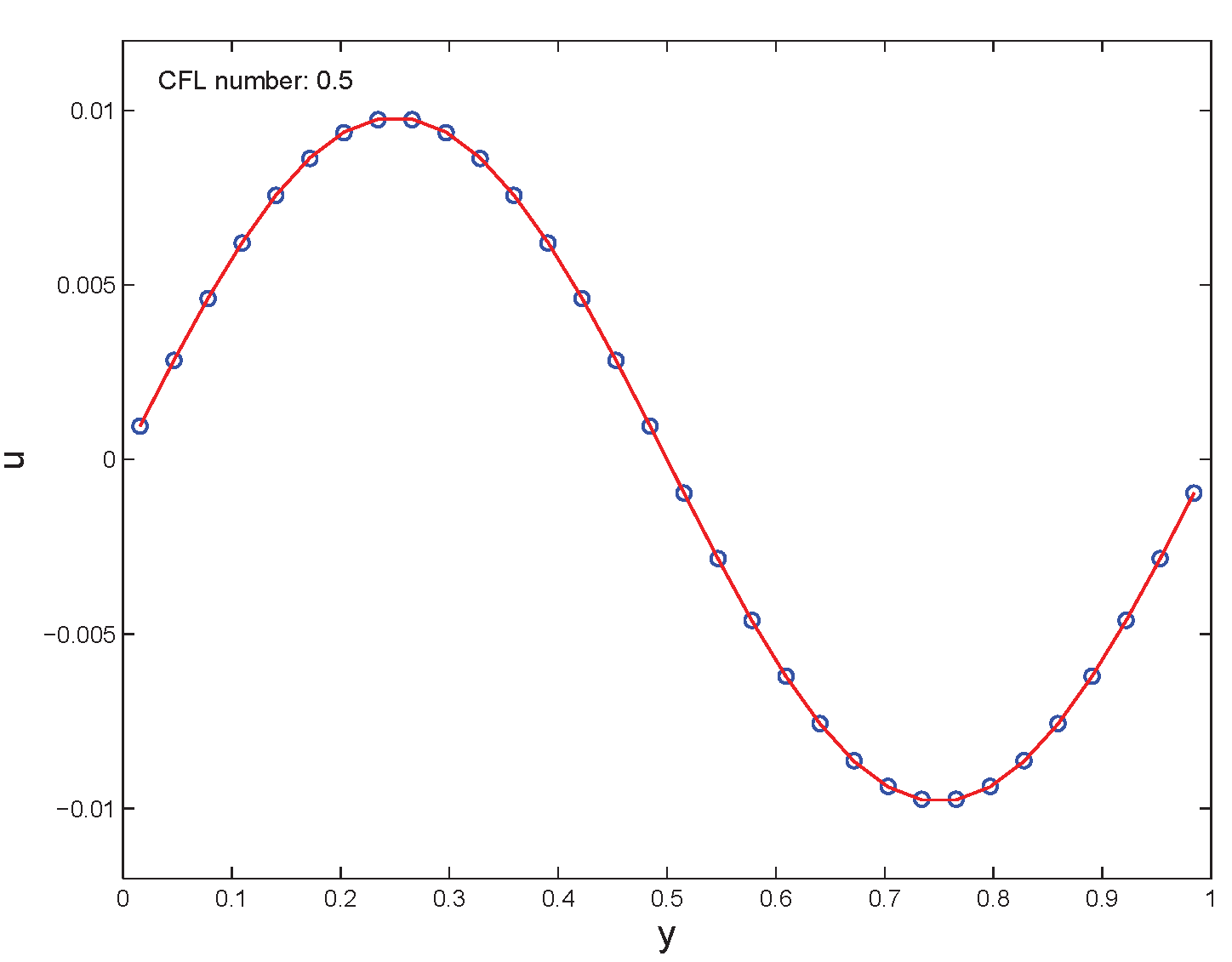}}
\subfigure[]{
\label{fig:PF_pressure:0.5}
\includegraphics[width=2.7in,height=2.3in]{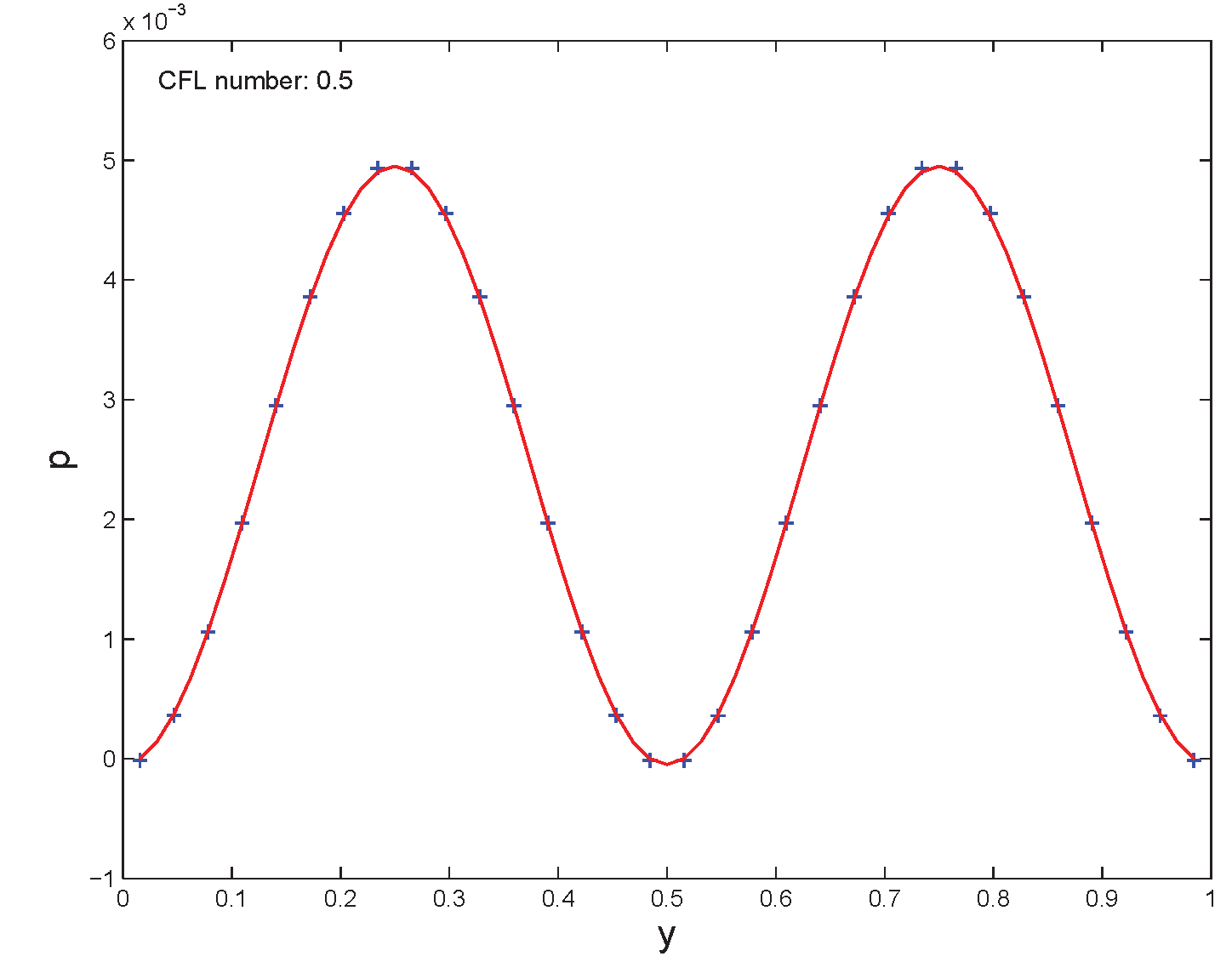}}
\subfigure[]{
\label{fig:PF_velocity:0.9}
\includegraphics[width=2.7in,height=2.3in]{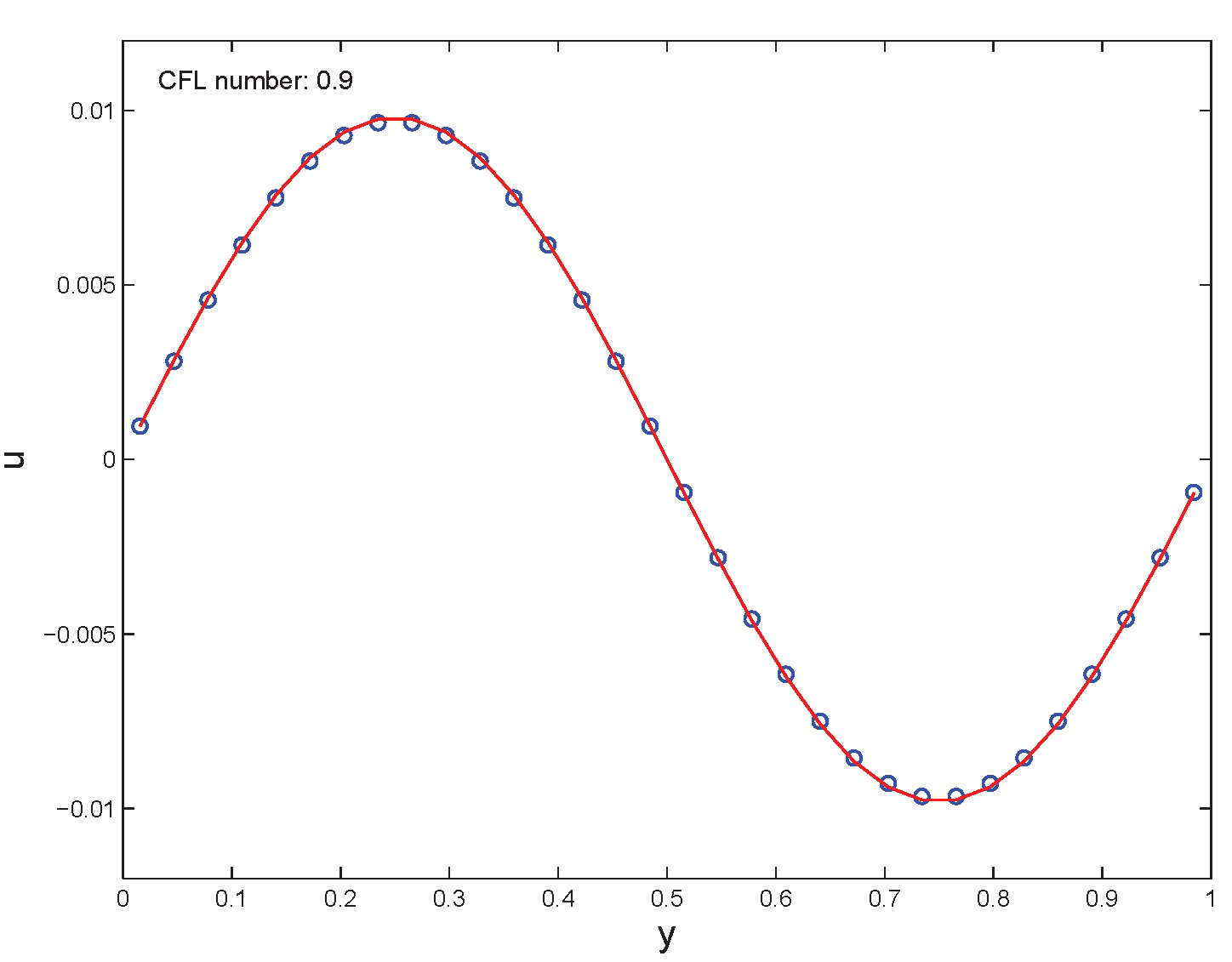}}
\subfigure[]{
\label{fig:PF_pressure:0.9}
\includegraphics[width=2.7in,height=2.3in]{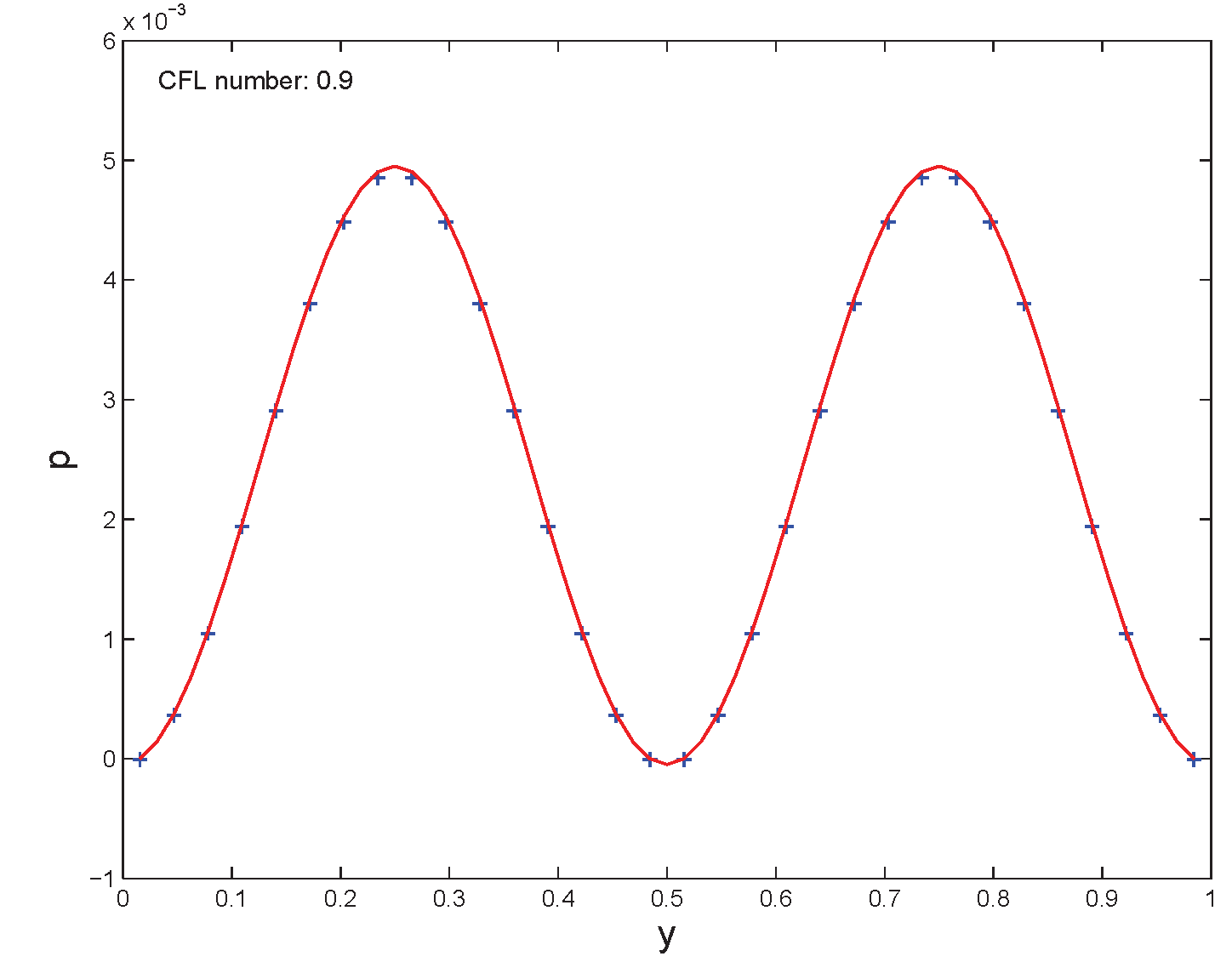}}
\caption{The velocity ($\circ$) and pressure ($+$) profiles of $\bm{u}_x(0.5, y)$ and $p(0.5, y)$ for the periodic flow with different CFL numbers, where the solid lines are analytical solutions.} 
\label{fig:PF_velocity}
\end{figure}

In our simulations, the parameters are set as follows: $ Re = 10$, $u_0 = 0.1$, $RT = 5$, $\rho_0 = 1.0$, and the CFL numbers are $0.3, 0.5$ and $0.9$, where the Reynolds number is defined as $Re = {u_0 L}/{\nu}$. The equilibrium distribution is initialized by using the analytical solution of velocity $u$ and using $\rho = \rho_0 + p/RT$ with the pressure $p$ given by Eq. \eqref{PF_AS}. To determine whether the steady state is reached, the following criterion is used:

\[\frac{{\sqrt {\sum\nolimits_{i,j} {{{\left| {u_{ij}^{(n + 1000)} - u_{ij}^{(n)}} \right|}^2}} } }}{{\sqrt {\sum\nolimits_{i,j} {{{\left| {u_{ij}^{(n + 1000)}} \right|}^2}} } }} \le {10^{ - 6}},\]
where $u_{ij}^n = u\left( {{x_i},{y_j},n\Delta t} \right)$. In Fig. \ref{fig:PF_velocity}, the velocity and pressure profiles computed from analytical solution and predicted by the incompressible DUGKS model on a $32 \times 32$ uniform mesh are both presented. It shows that the numerical results with different CFL numbers are all in excellent agreement with the analytical ones, which also implies that the proper CFL numbers have little influence on the proposed model.

In order to analyze the convergence order of the proposed incompressible DUGKS model, different meshes ($ N \times N, N = 16, 32, 64, 128$) are used for our simulations. To reduce the time error in the evaluation, we set the time step $\Delta t$ as a small value ($10^{-4}$). Then, we measured the $L_2$ relative global errors of steady velocity and pressure field:
\begin{equation}
\begin{split}
E(\phi ) &= \frac{{\sqrt {\sum\nolimits_{i,j} {{{\left| {\phi_{ij} - \phi '_{ij}} \right|}^2}} } }}{{\sqrt {\sum\nolimits_{i,j} {{{\left| {\phi '_{ij}} \right|}^2}} } }}, \\
\phi &= \bm{u} \ or \ p,
\end{split}
\end{equation}
where $\phi_{ij} = \phi(x_i, y_j)$, and $\phi_{ij} '$ is the analytical solution given by Eq. \eqref{PF_AS}.
The results in Table \ref{table:EC_VP} are clearly shown that the proposed incompressible DUGKS model is of second-order accuracy.

For this steady flow, we also test the compressible effect of both DUGKS and incompressible DUGKS model. The parameters are set as: $\rho_0 = 1.0$, viscosity $\nu = 10^{-2}$, $\Delta t = 10^{-4}$, the CFL number is 0.1, and the mesh is $32 \times 32$ uniform mesh. A series of simulations for different Mach numbers $Ma$ have been done, and the $L_2$ errors of velocity are measured.

\begin{figure}
\centering
\includegraphics[width=2.7in]{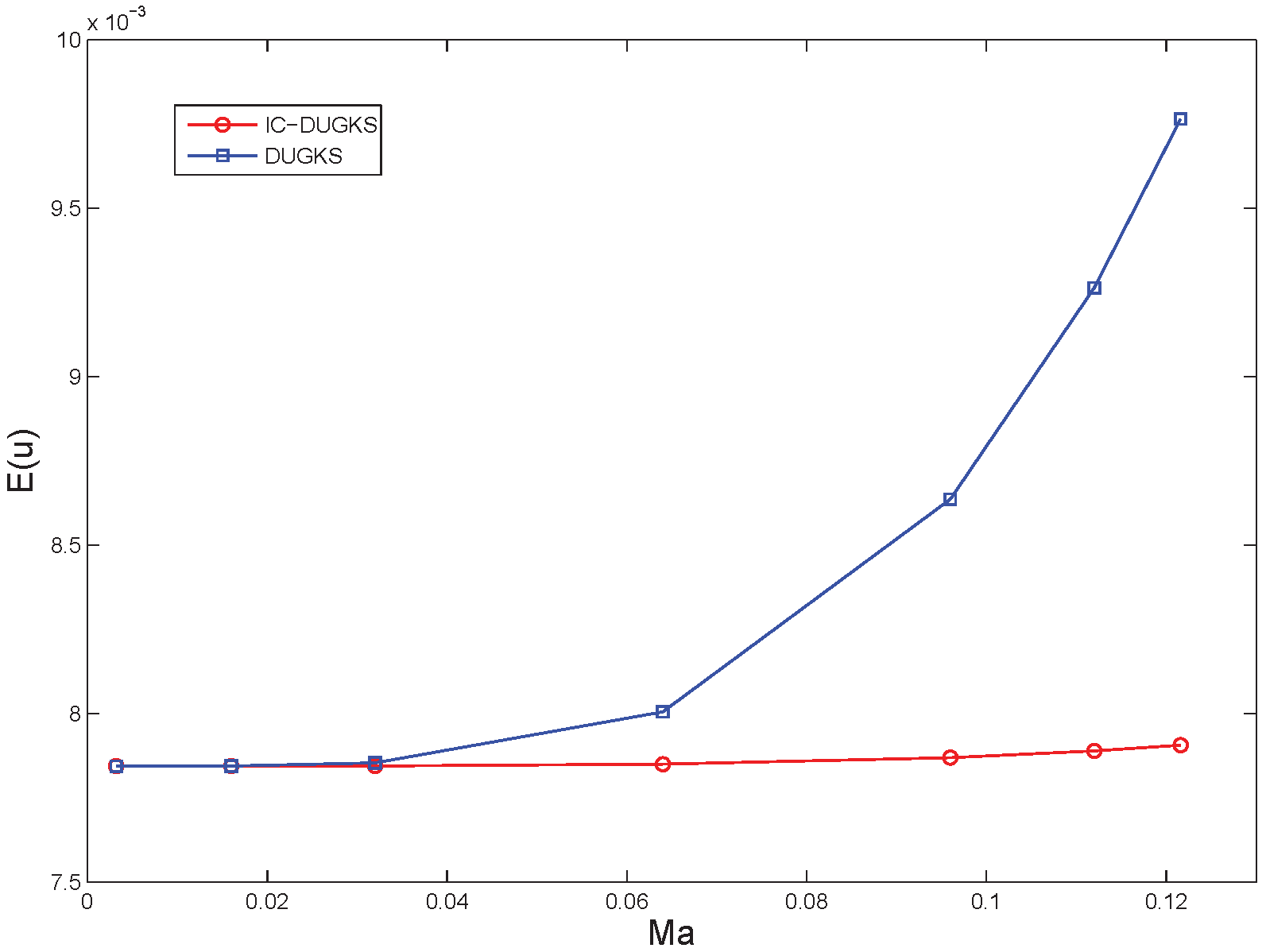}
\caption{$L_2$ errors of velocity measured from incompressible DUGKS and DUGKS for different Mach number.}
\label{fig:ICD_D_Ma}
\end{figure}

As shown in Fig. \ref{fig:ICD_D_Ma}, it is observed to find that at the beginning, the errors of DUGKS $E_c$ are almost the same with those of incompressible DUGKS $E_{ic}$, however, with the Mach number increasing, the growth of $E_c$ is significant with about $24.5\%$ from the beginning to the end, and yet, the growth of $E_{ic}$ is very small with only about $0.8\%$. Comparing with $E_c$ and $E_{ic}$ at the end point $Ma = 0.1216$, we can discover that $E_{ic}$ is about $19.0\%$ smaller than $E_c$ which indicates that the incompressible DUGKS model can reduce the compressible error efficiently. In general, for steady flow, the proposed incompressible DUGKS model is much less sensitive to Mach number than the original DUGKS model, and it also performs better in terms of accuracy with the Mach number increasing which indicates that the reduction of compressible errors is significant.

\begin{center}
\begin{table}
\caption{Errors and convergence orders of steady velocity and pressure.}
\begin{tabularx}{\textwidth}{l X<{\centering} X<{\centering} X<{\centering} X<{\centering}}
\hline
\hline
$N$          & $16$ & $32$    & $64$  & $128$ \\ 
\hline
$E(\bm{u})$  & $9.740 \times 10^{-3}$ & $2.410 \times 10^{-3}$ & $5.969 \times 10^{-4}$ & $1.446 \times 10^{-4}$ \\
 order       & $-$  & $2.015$ & $2.014$  & $2.025$   \\
$E(\bm{p})$  & $3.020 \times 10^{-2}$ & $7.412 \times 10^{-3}$ & $1.930 \times 10^{-3}$ & $5.840 \times 10^{-4}$ \\
 order       & $-$  & $2.027$ & $1.984$  & $1.898$   \\
\hline
\hline
\end{tabularx}
\label{table:EC_VP}
\end{table}
\end{center}

\subsection{Plane Poiseuille flow}
\label{sec:Poiseuille}
In this example, we will test the bounce-back (BB) method and the non-equilibrium extrapolation (NEQ) method. 

The test case is the steady plane Poiseuille flow driven by a force $\rho \bm{G}$ which is defined in the region $\left\{ {(x,y)|0 \le x \le 1,0 \le y \le 1} \right\}$ with periodic boundary condition at the entrance and exit and using the BB and NEQ methods at both the solid tube walls. The analytical solution is given by,
\begin{equation}\label{Poiseuille_AS}
\begin{split}
u(x,y) &= \frac{G}{{2\nu }}y(1 - y),\\
v(x,y) &= 0.
\end{split}
\end{equation}

\begin{figure}
\centering
\includegraphics[width=2.7in]{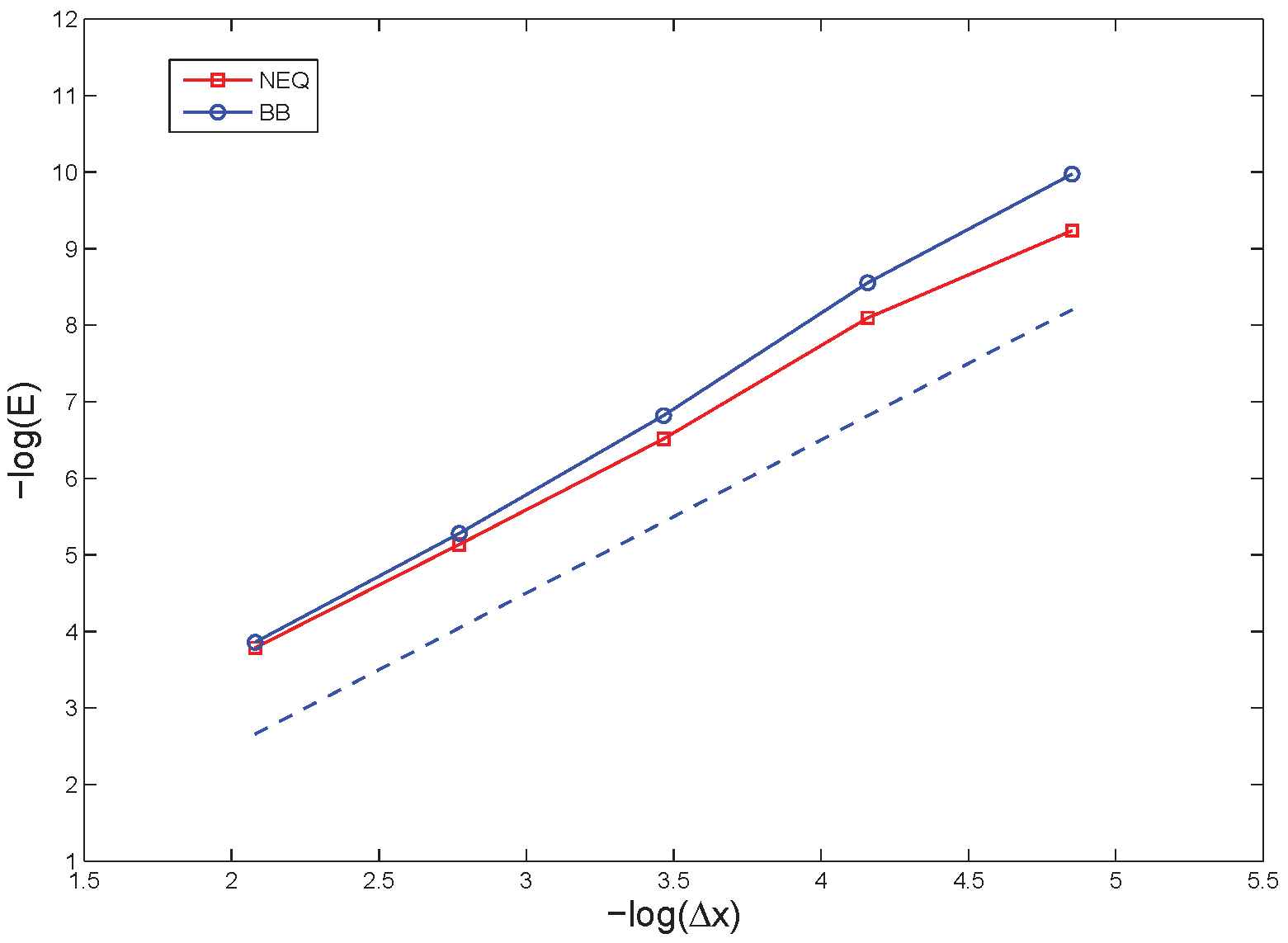}
\caption{Relative errors of velocity ($E$) vs. space steps($\Delta x$). The slope of dash line is $2$.}
\label{fig:Boundary_order}
\end{figure}

To analyze the convergence order of th BB and NEQ methods, five different meshes ($ N \times N, N = 8, 16, 32, 64, 128$) are employed for our simulations, and other parameters are set as follows, $\rho_0 = 1.0$, $Re = 10$, $G = 10^{-3}$, $\Delta t = 10^{-4}$, where the Reynolds number is defined as $Re = u_{max} L / \nu$ and $u_{max} = L^2G/8\nu$. The equilibrium distribution is initialized from the analytical solution of velocity given by Eq. \eqref{Poiseuille_AS} and $\rho_0$. The steady criterion and the $L_2$ relative errors are defined as the same in Section \ref{sec:Periodic_flow}. As the results shown in Fig. \ref{fig:Boundary_order}, it appears that both BB and NEQ methods can achieve second-order. Even though the BB method performs better in accuracy, the errors of the two methods are very close for the maximum difference of errors between two methods is $0.16\%$ (on the mesh of $8 \times 8$). Evidently, the non-equilibrium extrapolation method which is also widely used in LBE is suitable for DUGKS model as well as the bounce-back method. In next section, the NEQ method will be used to process the pressure boundary condition which the bounce-back method cannot do.

\subsection{Womersley flow}
\label{sec:Womersley}
To validate the proposed model for unsteady flow, the 2D Womersley flow driven by periodic pressure gradient is employed. With the pressure gradient $\partial P/\partial x = G\cos (\omega t)$, the analytical solution of velocity is given by,
\begin{equation}{\label{Wom_AS_Complex}}
{u_x}(y,t) = \text{Re}\left[ {i\frac{G}{\omega}\left( {1 - \frac{{\cos \left[ {\lambda (2y/{L_y} - 1)} \right]}}{{\cos \lambda }}} \right){e^{i\omega t}}} \right],
\end{equation}
where $G$ is the amplitude of the varying pressure gradient, $\lambda$ and the Womersley number $\alpha$ is defined as
\begin{equation}
\lambda  =  - i{\alpha ^2}, \ \ a = \frac{{{L_y}\omega }}{{4\nu }},
\end{equation}
Besides, if we set the pressure at exit to be a constant $p_{out}$, then we can obtain the analytical solution of pressure, as
\begin{equation}{\label{Wom_AS_Pressure}}
p(x,t) = {p_{out}} - ({L_x} - x)G\cos \omega t.
\end{equation}

In addition, the analytical solution of velocity also can be rewritten as,
\begin{equation}{\label{Wom_AS_Velocity}}
{u_x}(y,t) = \frac{G}{\omega }\left[ { - \sin \omega t + \frac{{{K_r}\sin \omega t}}{{{K_a}}} + \frac{{{K_i}\cos \omega t}}{{{K_a}}}} \right],
\end{equation}
where
\begin{equation}
\begin{split}
K_r=\cos (\theta )\cosh(\theta )\cos(k)\cosh(k) &+ \sin (\theta )\sinh (\theta )\sin (k)\sinh (k), \\
K_i=\sin (\theta )\sinh (\theta )\cos(k)\cosh(k) &- \cos (\theta )\cosh(\theta )\sin (k)\sinh (k), \\
K_a={\left[ {\sin (\theta )\sinh (\theta )} \right]^2} &+ {\left[ {\cos (\theta )\cosh(\theta )} \right]^2}, \\
k=\sqrt{2}\alpha/2, \ \ \theta&=k(2y/L_y-1).
\end{split}
\end{equation}

\begin{figure}
\subfigure[]{
\label{fig:Wom_velocity}
\includegraphics[width=2.7in,height=2.3in]{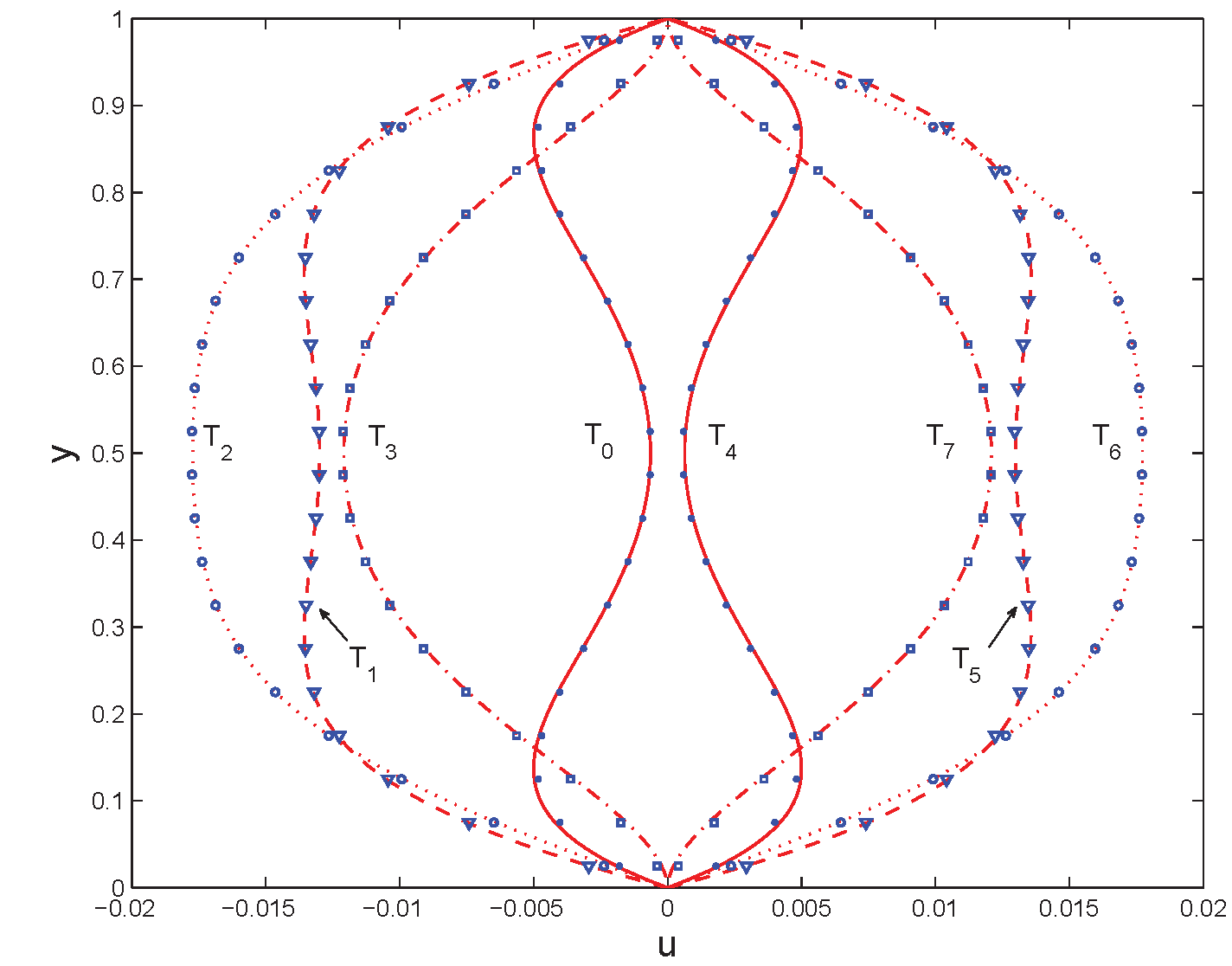}}
\subfigure[]{
\label{fig:Wom_pressure}
\includegraphics[width=2.7in,height=2.3in]{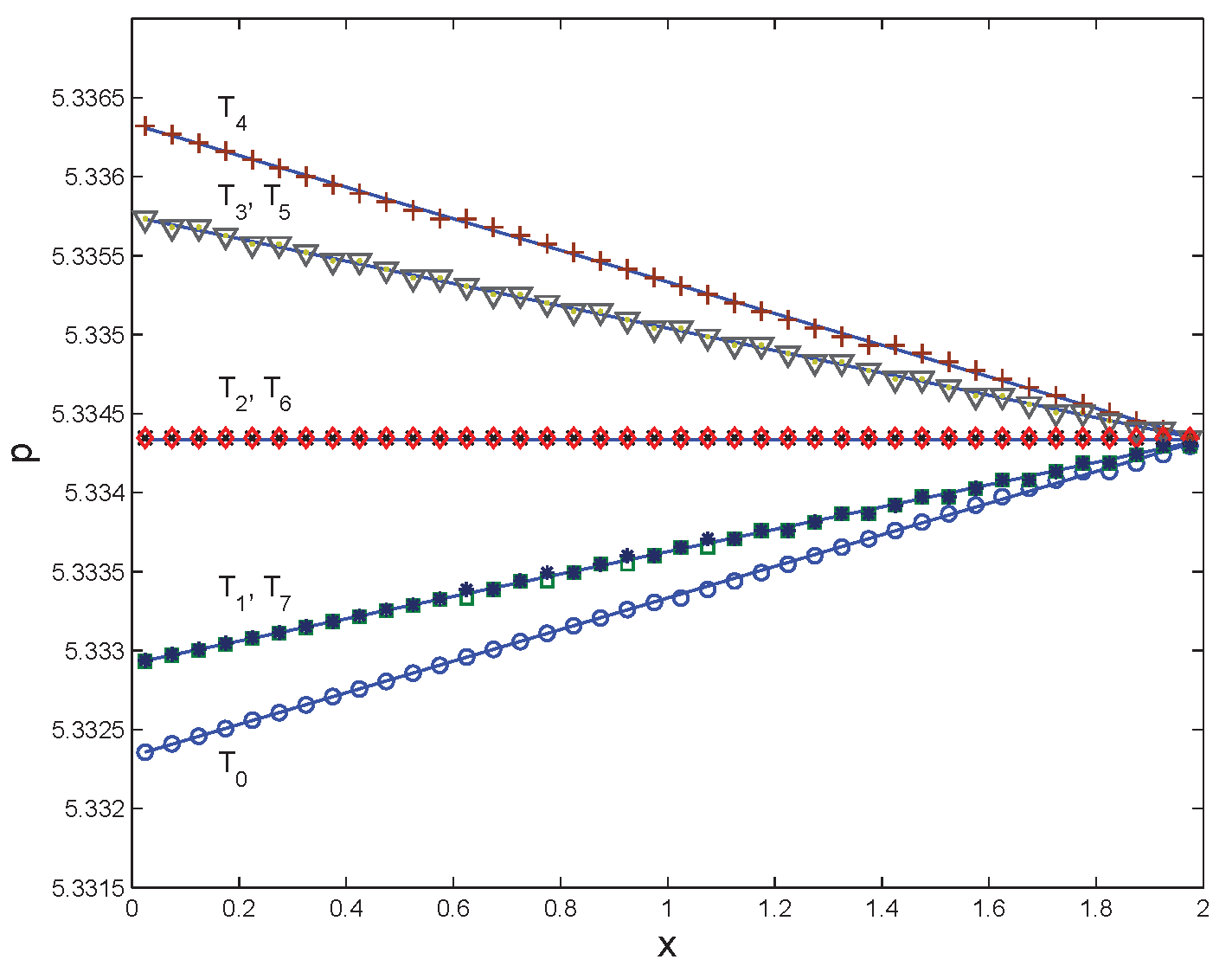}}
\caption{The velocity and pressure profiles of $\bm{u}_x(1.0, y, t)$ and $p(x, 0.5, t)$ for the Womersley flow at different times ($t = T_n \stackrel{\Delta}{=} nT/8, n = 0,1,\ldots,7$), where the markers are numerical results and the lines are analytical solutions.} 
\label{fig:Wom}
\end{figure}

In our simulations, the Womersley flow is defined in the region $\left\{ {(x,y)|0 \le x \le 2,0 \le y \le 1} \right\}$ with pressure boundary condition at the entrance and exit using the NEQ method, and with solid tube walls at up and down using the BB methods. The parameters are set as follows: the mesh is $40 \times 20$, $\rho_0=1.0$, $\nu = 0.001$, $G=0.001$ ($\Delta P = L_xG\cos\omega t$), the CFL number is 0.5, $RT=16/3$, the period of the driven pressure is $T=100$ and $\omega=2\pi /T$, then the Womersley number should be $\alpha=3.963$. The equilibrium distribution is initialized from the analytical solution of pressure and velocity given by Eq. \eqref{Wom_AS_Pressure} and Eq. \eqref{Wom_AS_Velocity}. The criterion of convergence is defined by
\[\frac{{\sqrt {\sum\nolimits_{i,j} {{{\left| {u_{ij}^{n + T} - u_{ij}^{n}} \right|}^2}} } }}{{\sqrt {\sum\nolimits_{i,j} {{{\left| {u_{ij}^{n + T}} \right|}^2}} } }} \le {10^{ - 8}},\]
where $u_{ij}^n = u\left( {{x_i},{y_j},n\Delta t} \right)$. After the system reached the convergence criterion, the velocity and pressure profiles are measured at eight different times ($t = nT/8, n=0,1,\ldots,7$). As shown in Fig. \ref{fig:Wom}, the numerical results of velocity are in good agreement with the analytical solutions given by Eq. \eqref{Wom_AS_Velocity}, and the pressure ones have little fluctuations along the analytical results given by Eq. \eqref{Wom_AS_Pressure}, but the $L_2$ relative global errors of pressure over the period $T$ are all less than $0.42\%$. In general, the numerical results of velocity and pressure are both agreed well with the analytical solutions.

To compare the incompressible DUGKS model with the original model, we conduct a set of simulations with increasing amplitude of the pressure gradient $G$ (or the maximum Mach number $M_{max} = U_{max}/ C$ and $U_{max}$ is the maximum velocity appearing in the tube axis) and measured the $L_2$ relative global errors in the velocity fields, which is defined by
\begin{equation}
E(u) = \frac{{\sqrt {\sum\nolimits_{i,j} {{{\left| {u_{ij}^{n} - \bar u_{ij}^{n}} \right|}^2}} } }}{{\sqrt {\sum\nolimits_{i,j} {{{\left| {\bar u_{ij}^{n}} \right|}^2}} } }}, 
\end{equation}
where $\bar u$ is the analytical solution given by Eq. \eqref{Wom_AS_Velocity}. In Table. \ref{table:Wom_C_IC}, the maximum error $E_{max}$ and the average error $<E>$ are presented, where the $E_{max}$ is maximum value of $E$ in the period $T$ and the $<E>$ is averaged over a period ($t = nT/8, n=0,1,\ldots,7$). As the results show, with the increasing $M_{max}$ (or $G$), the corresponding errors of the original model grow faster than those of the incompressible DUGKS model. Besides, the accuracy of the incompressible DUGKS model is better than the original one especially with a higher Mach number. Obviously, the proposed model can also reduce the compressible error efficiently for the unsteady flow.

\begin{center}
\begin{table}
\caption{Maximum and average errors of velocity in Womersley flow, $\alpha = 1.253$}

\begin{tabularx}{\textwidth}{X<{\centering} X<{\centering} X<{\centering} X<{\centering} l X<{\centering} X<{\centering}}
\hline
\hline
\multirow{2}{*}{} & & \multicolumn{2}{c}{Original model} & \ & \multicolumn{2}{c}{Present model} \\
\cline{3-4}
\cline{6-7}
$G$  & $M_{max}$ & $E_{max}$    & $<E>$ & \ & $E_{max}$  & $<E>$\\ 
\hline
$0.005$ & $0.0129$ & $2.29 \%$  & $0.87 \%$  & \ & $2.15 \%$ & $0.87 \%$ \\
$0.01$  & $0.0258$ & $2.54 \%$  & $0.88 \%$  & \ & $2.30 \%$ & $0.87 \%$ \\
$0.05$  & $0.1289$  & $4.31 \%$  & $2.63 \%$  & \ & $3.42 \%$ & $1.04 \%$ \\
$0.1$   & $0.2578$  & $12.18 \%$ & $7.63 \%$  & \ & $4.71 \%$ & $1.70 \%$ \\
$0.15$  & $0.3867$  & $21.03 \%$ & $13.70 \%$ & \ & $5.95 \%$ & $2.46 \%$ \\
\hline
\hline
\end{tabularx}
\label{table:Wom_C_IC}
\end{table}
\end{center}

\subsection{2D lid-driven flow}
\label{sec:2D_LDF}
The classical 2D lid-driven flow (LDF) is a standard benchmark problem that has been investigated by many authors \cite{ladd1994numerical, ghia1982high, cazemier1998proper, bruneau20062d}. In this section, we simulate this problem which has no analytical solution for testing the proposed model. The configuration considered is a 2D square cavity with a top wall moving with a constant velocity $U$ along the horizontal direction and the other three walls are fixed. The Reynold number is defined as $Re = UL/\nu$, where $L$ is the cavity length.

By using the proposed model, numerical simulations are carried out for the lid-driven flow at different Reynold numbers $Re = 400, 1000, 5000, 7500$ on the $80 \times 80$ uniform mesh. The driven velocity is set to be $U=1.0$ and $RT = 100/3$ so that the small Mach number can be promised, and the NEQ method is used to treat the velocity boundary condition. The length of cavity is $L=1.0$ and the CFL number is set to be $0.5$. The equilibrium distribution is initialized by $u_x = 0$, $u_y = 0$ and $\rho = 1$. We also employ the criterion of convergence which is defined in Sec. \ref{sec:Periodic_flow} to this problem.

\begin{figure}
\subfigure[]{
\label{fig:LDF_u}
\includegraphics[width=6.3in,height=3in]{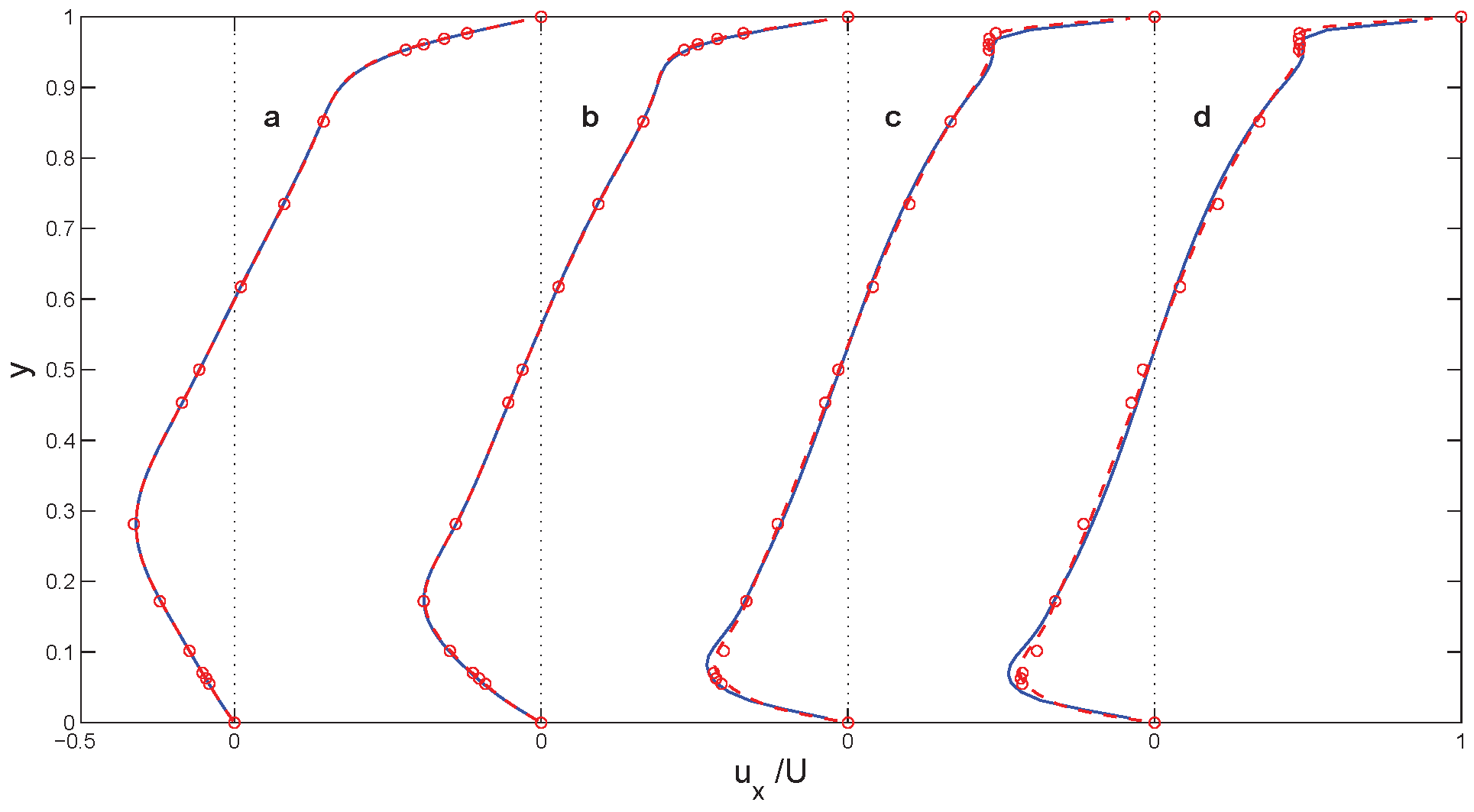}}
\subfigure[]{
\label{fig:LDF_v}
\includegraphics[width=6.3in,height=3in]{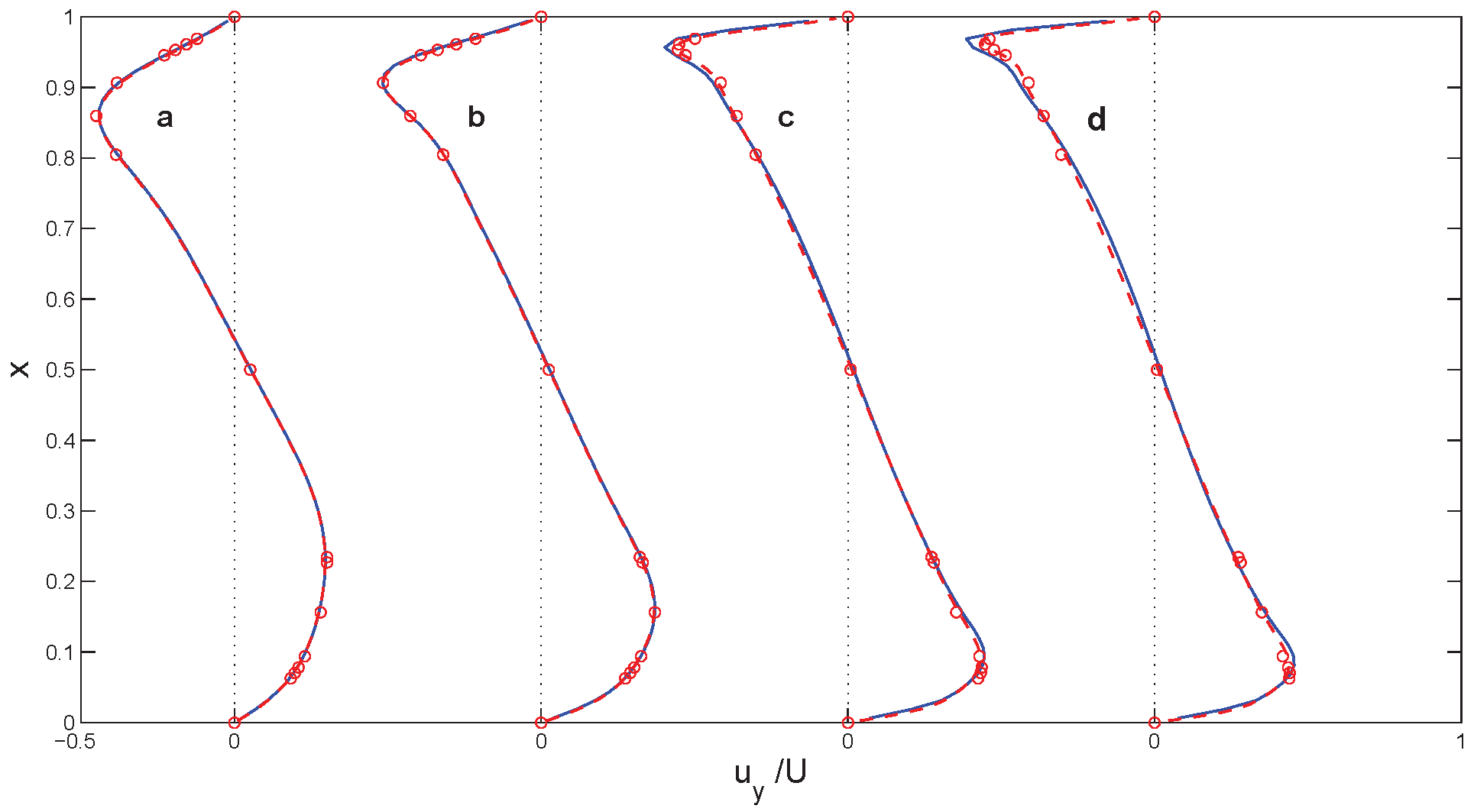}}
\caption{Comparison of velocity profiles with reference data \cite{ghia1982high} at four different Reynold numbers ($\bm{a}$. $Re=400$, $\bm{b}$. $Re=1000$, $\bm{c}$. $Re=5000$, $\bm{d}$. $Re=7500$), where the solid lines are numerical results on uniform mesh, the dashed lines are numerical results on nonuniform mesh, and ($\circ$) is reference data. (a) shows $u_x$ along the vertical line through the cavity center; (b) shows $u_y$ along the horizontal lines through the cavity center.} 
\label{fig:LDF_velocity}
\end{figure}

Firstly, we considered the profiles of the velocity component, $u_x$ and $u_y$, along vertical and horizontal center lines. As the solid lines shown in Fig. \ref{fig:LDF_velocity}, we can find that the numerical results are in good agreement with the reference data for $Re=400, 1000$. However, for higher Reynold numbers $Re=5000,7500$, it can be observed that although the numerical results still agree well with the reference data at middle of the region, the differences near the boundaries are obvious. This is because, when the Reynold number becomes higher, the flow field will be more complex especially at the boundaries and corners, and the current mesh appears not fine enough to describe the flow field, so it will cause some discrepancies.

\begin{figure}
\includegraphics[width=2.7in,height=2.7in]{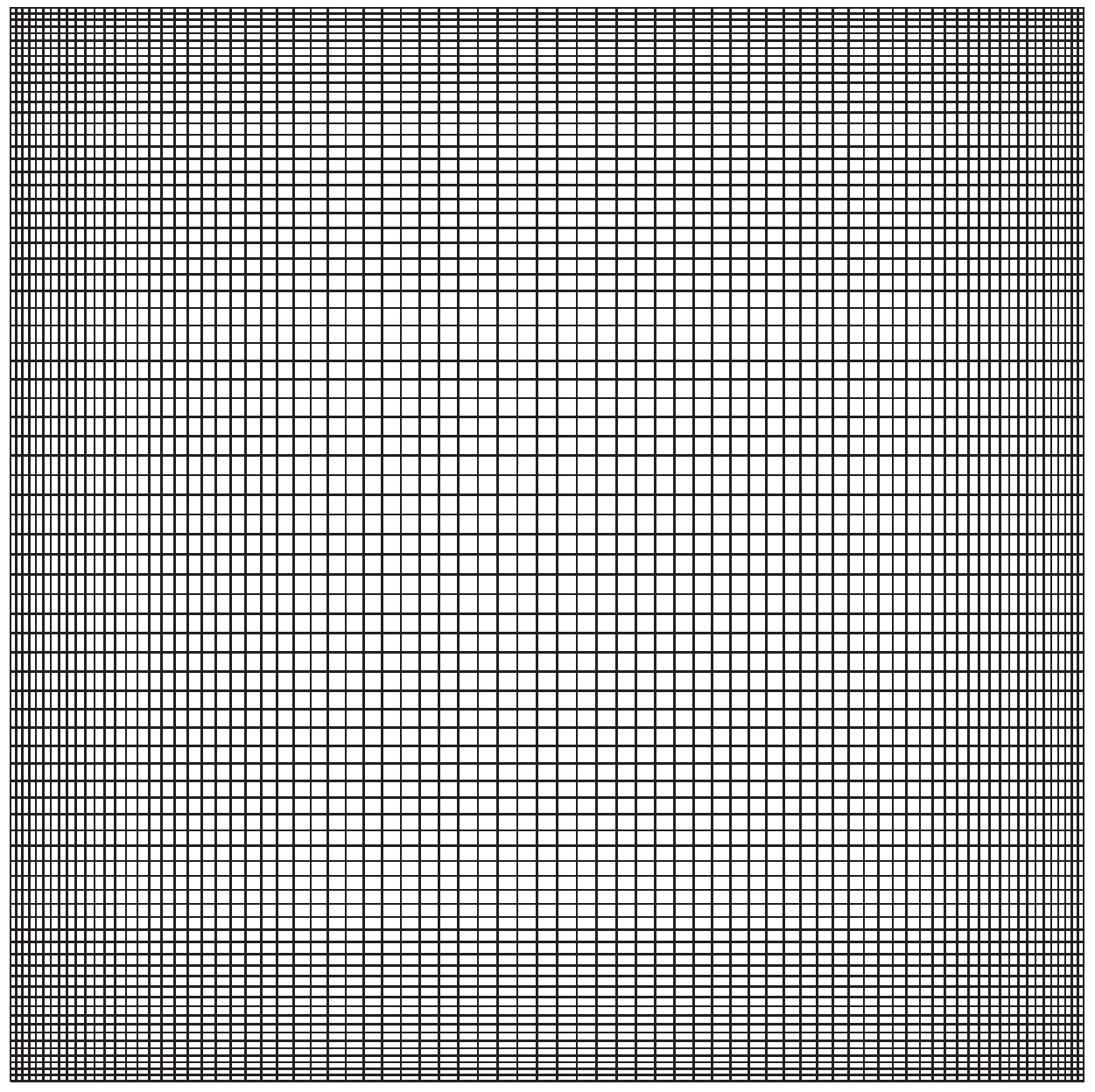}
\caption{$80 \times 80$ nonuniform mesh with $k=2.5$.} 
\label{fig:LDF_mesh}
\end{figure}

As a finite volume method, nonuniform mesh can also be applied for the proposed model. Comparing with the uniform mesh, we also simulate the lid-driven flow by using the $N \times N = 80 \times 80$ nonuniform mesh with locally refined meshes near the boundaries and corners. The mesh points $(x_i, y_j)$ are generated as follows,
\begin{equation}{\label{Nonuniform_mesh}}
\begin{split}
&{\zeta _i} = 0.5 + \frac{{\tanh [k(i/N - 0.5)]}}{{2\tanh (k/2)}}, \ i = 0,1,...,N, \\
&\left\{ {\begin{array}{*{20}{c}}
{{x_i} = \frac{{L({\zeta _i} + {\zeta _{i + 1}})}}{2}}\\
{{y_j} = \frac{{L({\zeta _j} + {\zeta _{j + 1}})}}{2}}
\end{array}} \right.,0 \le i,j \le N - 1,
\end{split}
\end{equation}
and the distribution of the grid is determined by the constant $k$. In current simulation, $k$ is set to be $2.5$, and the mesh is shown in Fig. \ref{fig:LDF_mesh}.

\begin{figure}
\subfigure[]{
\label{fig:LDF_PU}
\includegraphics[width=2.7in,height=2.3in]{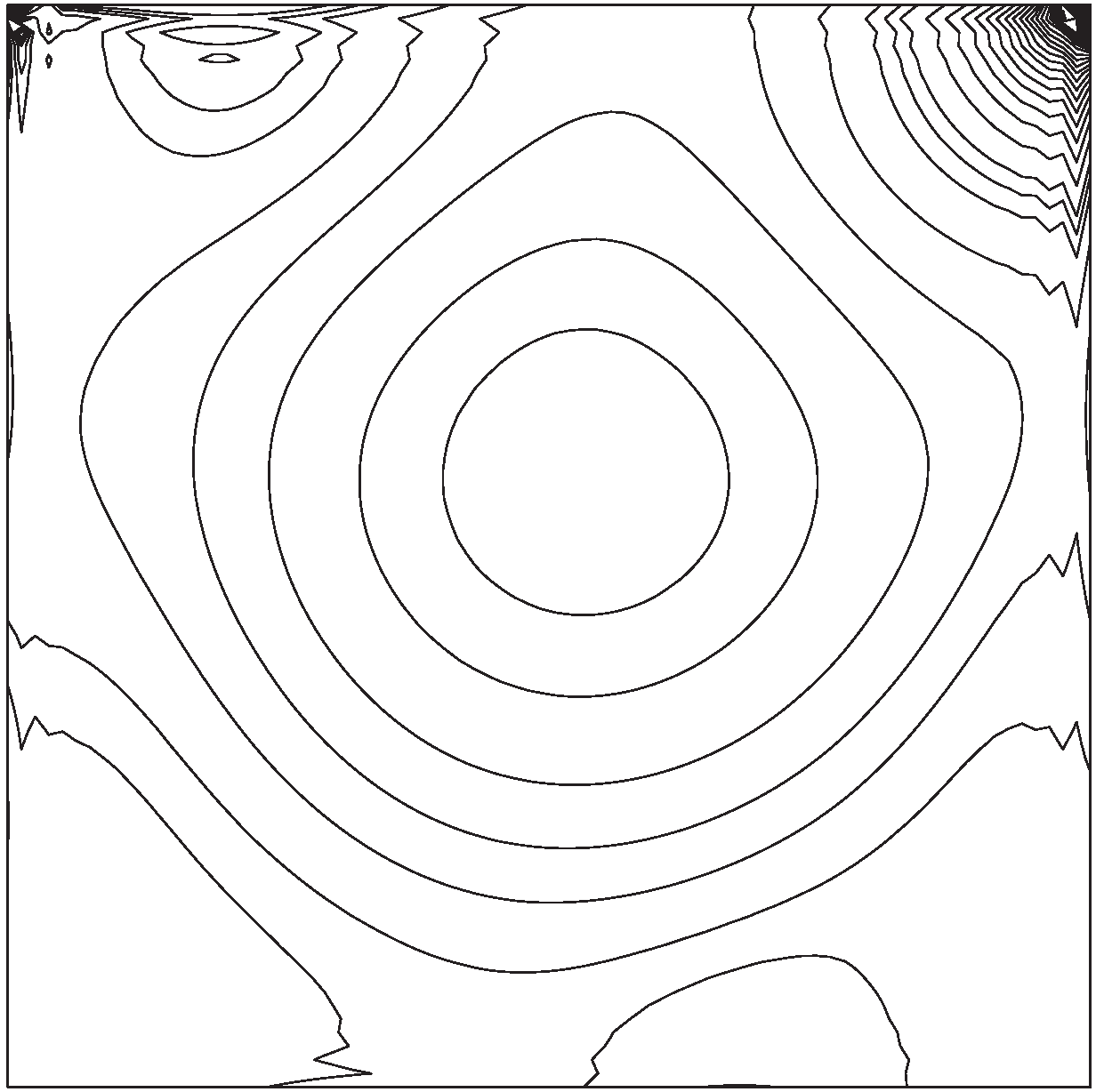}}
\subfigure[]{
\label{fig:LDF_PN}
\includegraphics[width=2.7in,height=2.3in]{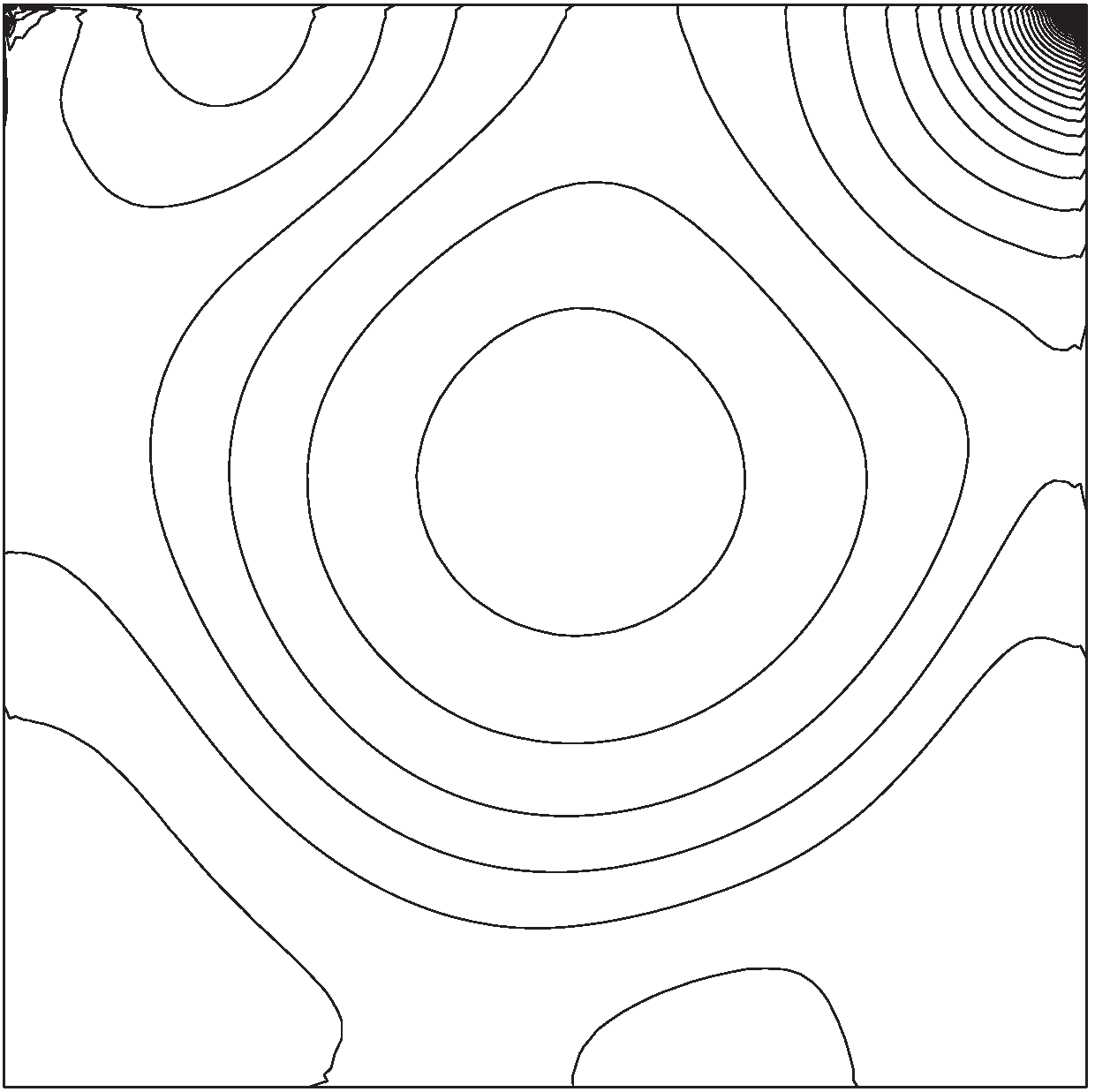}}
\caption{Pressure contours with uniform and nonuniform meshes at $Re=1000$.} 
\label{fig:LDF_P}
\end{figure}

Fig. \ref{fig:LDF_velocity} shows that both the results of uniform and nonuniform mesh have good agreement with the reference data at $Re=400,1000$. But, as the Reynold number increased to $Re = 5000, 7500$, the results of nonuniform mesh (dashed lines) perform more accurately than the uniform ones. In addition, as shown in Fig. \ref{fig:LDF_P}, we can clearly find that the pressure contour at $Re=1000$ computed by nonuniform mesh is better near the walls. Apparently, the proper nonuniform mesh can improve the prediction especially in the transition regions close to the boundaries.

We discussed the 2D-LDF at low Reynold numbers with steady solutions above, then we will validate the proposed model by simulating the 2D-LDF at high Reynold number ($Re = 12000$) whose solution is more complex and no longer steady \cite{cazemier1998proper}. The parameters for simulation are the same with above problems. The driven velocity is set to be $U=1.0$ and $RT = 100/3$, and the length of cavity is $L=1.0$ and the CFL number is set to be $0.5$.  The BB method is used to treat the velocity boundary condition. The equilibrium distribution is also initialized by $u_x = 0$, $u_y = 0$ and $\rho = 1$. In order to obtain accurate result, we use the $N \times N = 128 \times 128$ nonuniform mesh in this simulation as the Reynold number increased.

\begin{figure}
\subfigure[]{
\label{fig:LDF_hRe_KE}
\includegraphics[width=2.7in,height=2.3in]{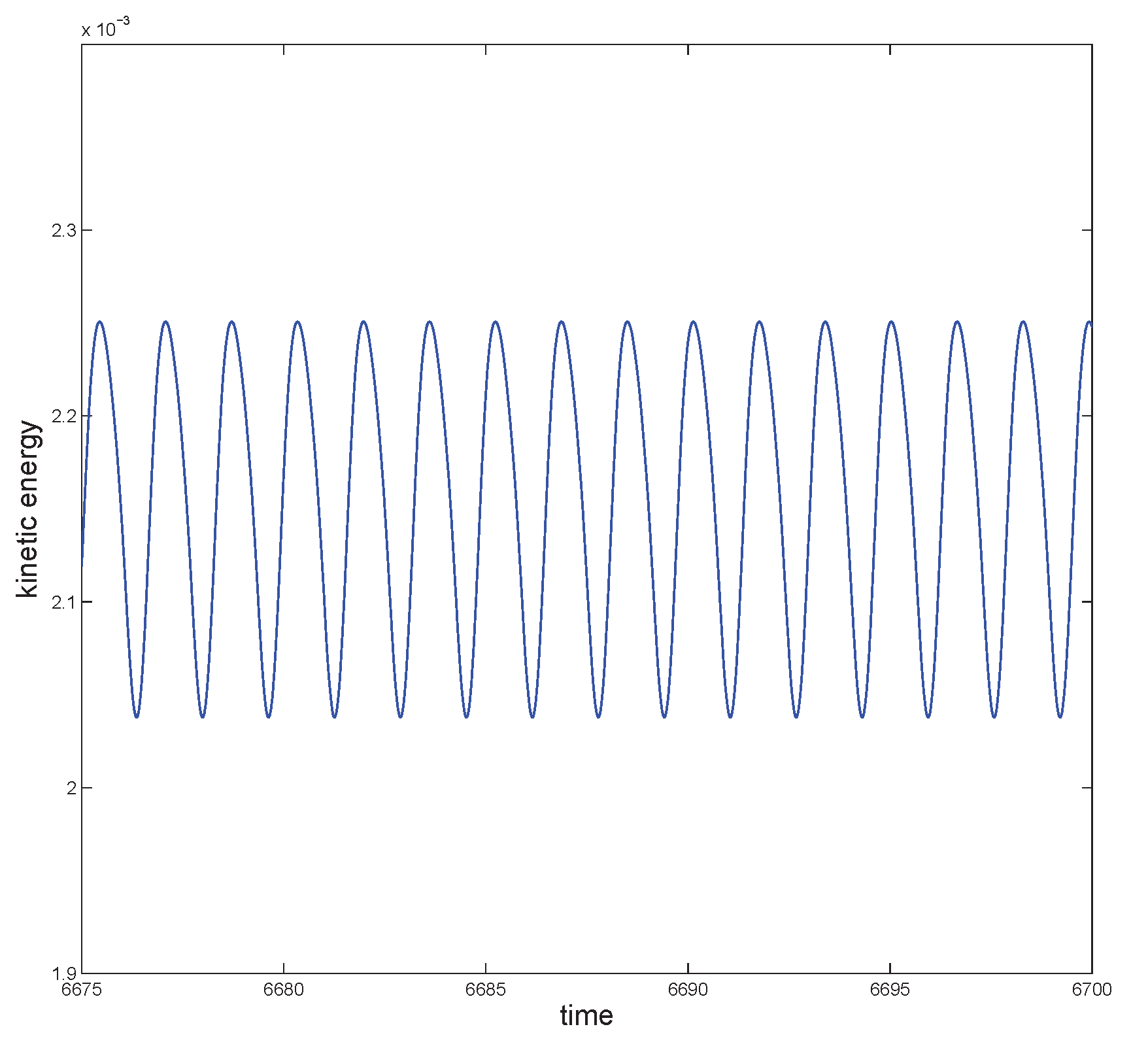}}
\subfigure[]{
\label{fig:LDF_hRe_KE_Freq}
\includegraphics[width=2.7in,height=2.3in]{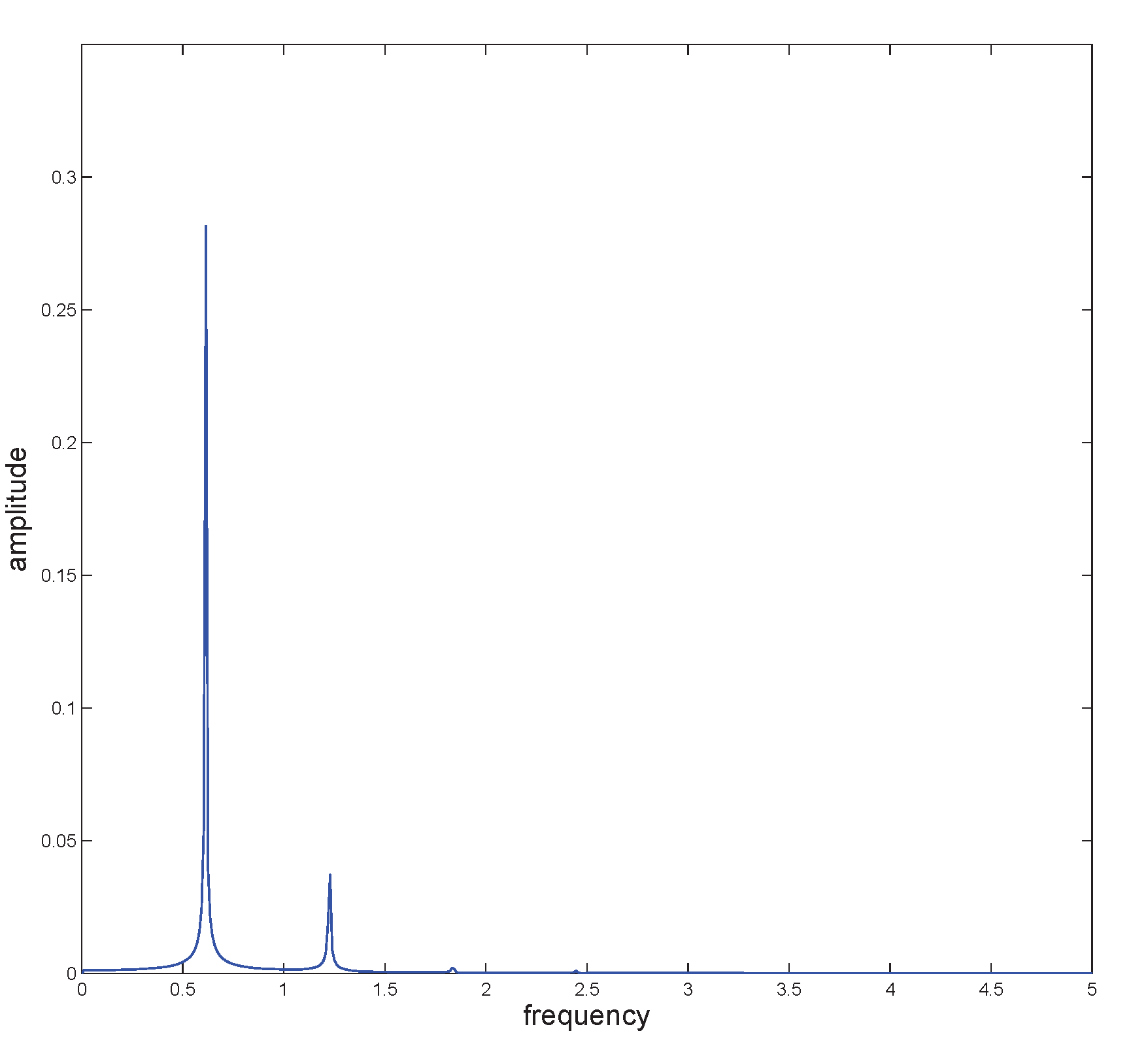}}
\caption{Kinetic energy history and its Fourier power spectrum for $Re=12000$.} 
\label{fig:LDF_hRe}
\end{figure}

\begin{figure}
\subfigure[]{
\label{fig:LDF_Period_Re12000_1}
\includegraphics[width=2.2in,height=1.9in]{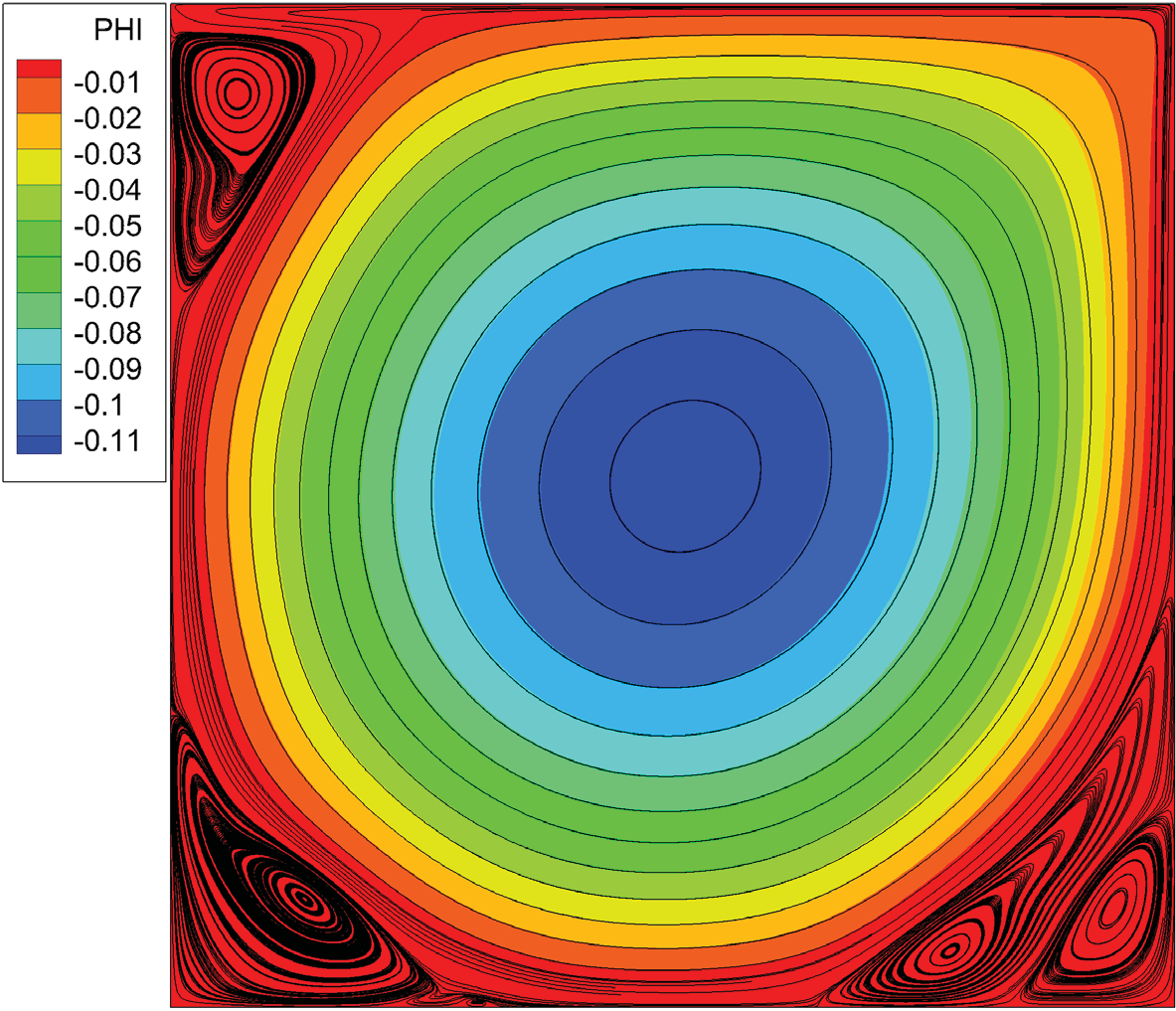}}
\subfigure[]{
\label{fig:LDF_Period_Re12000_2}
\includegraphics[width=1.9in,height=1.9in]{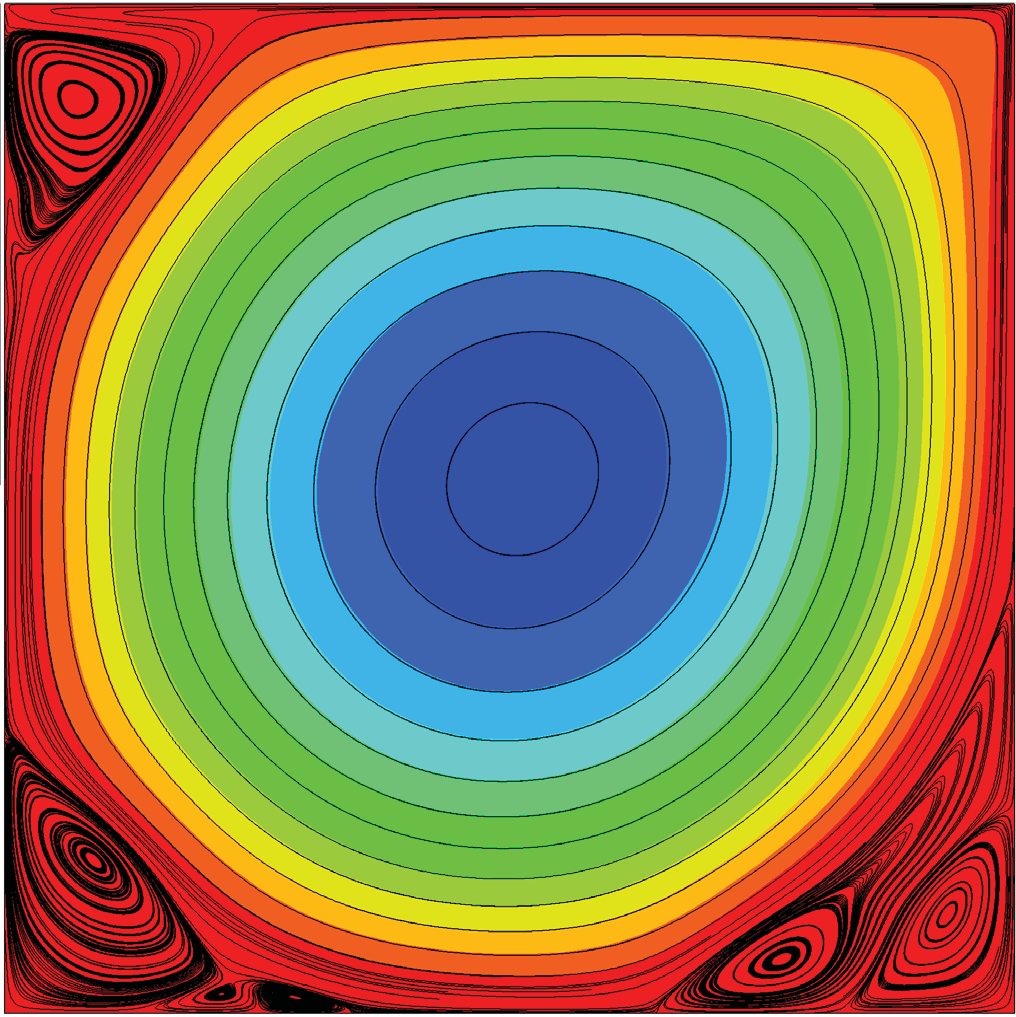}}
\subfigure[]{
\label{fig:LDF_Period_Re12000_3}
\includegraphics[width=1.9in,height=1.9in]{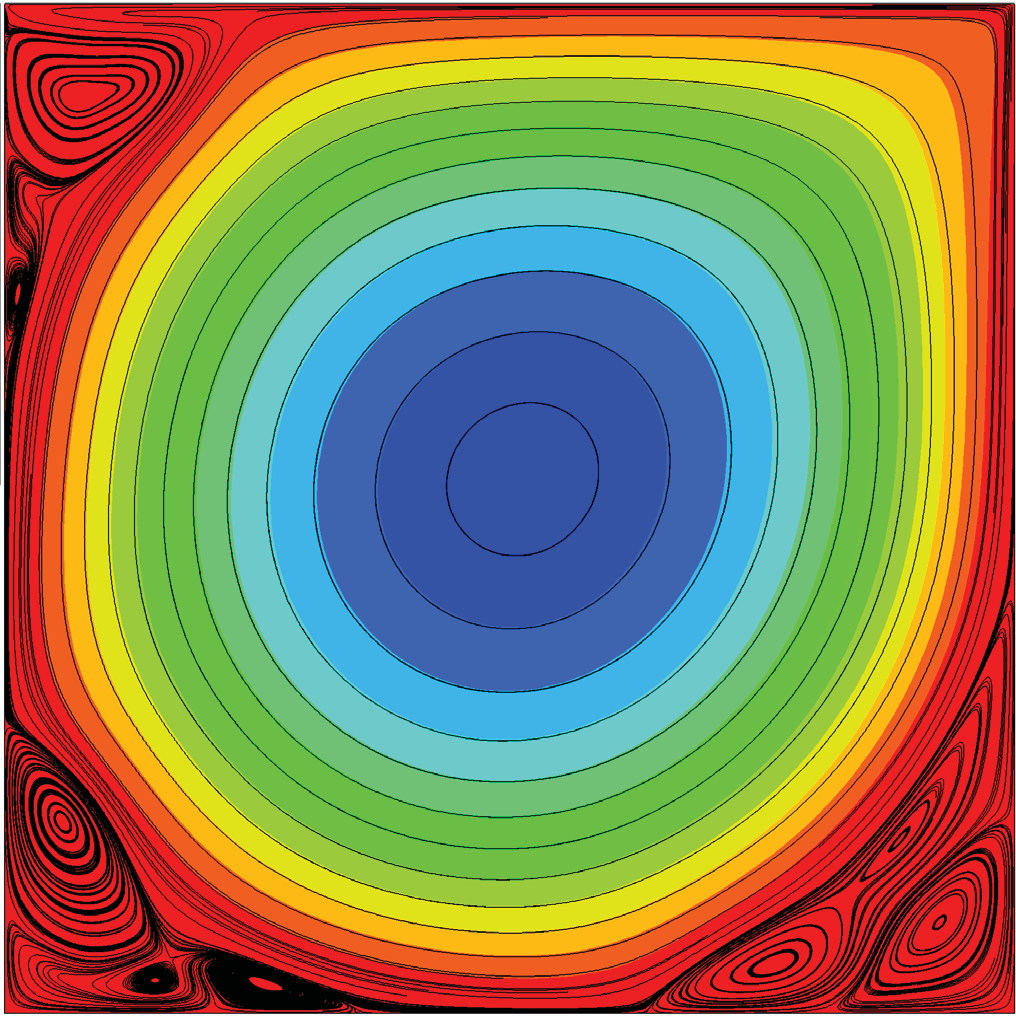}}
\subfigure[]{
\label{fig:LDF_Period_Re12000_4}
\includegraphics[width=2.2in,height=1.9in]{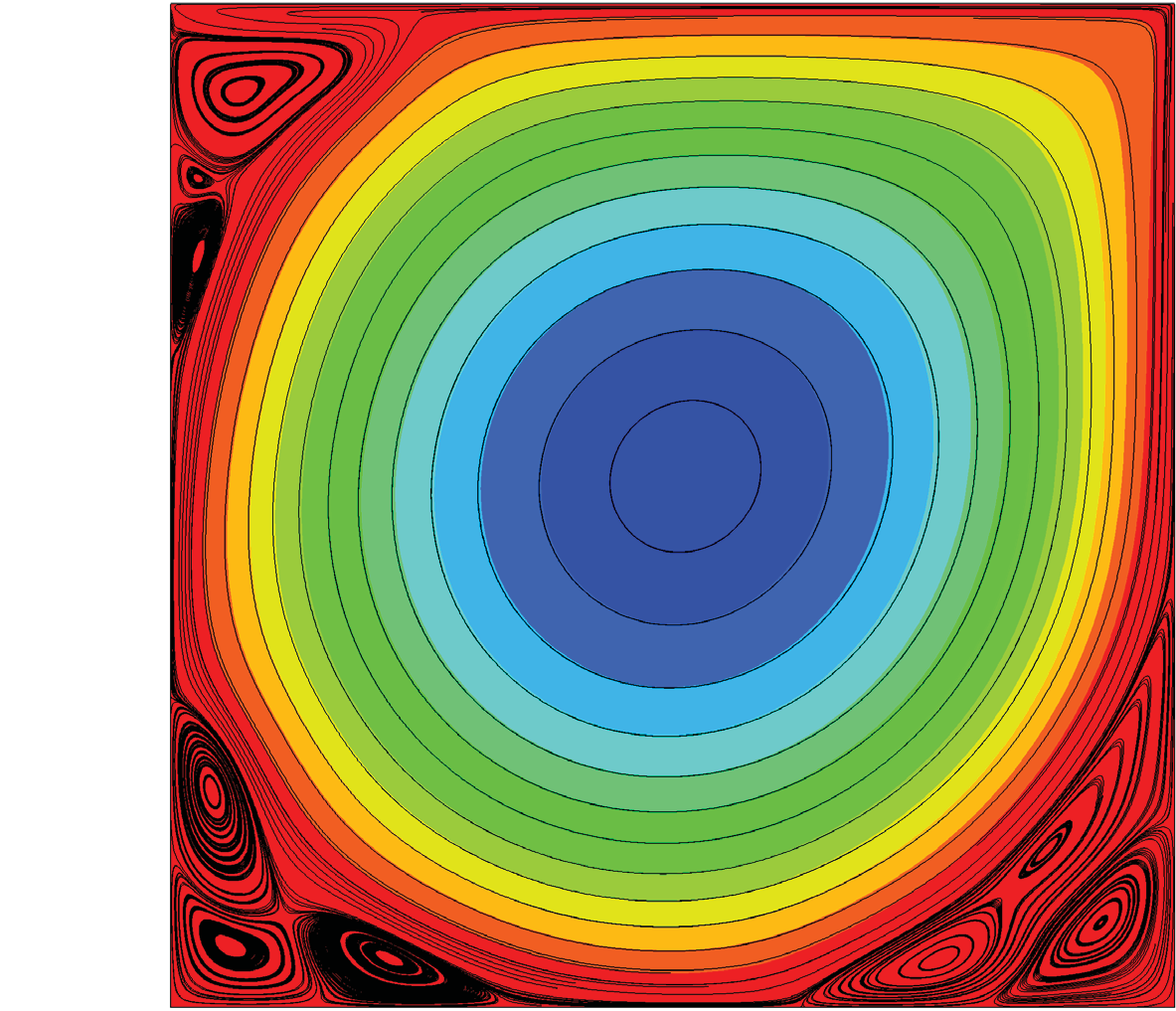}}
\subfigure[]{
\label{fig:LDF_Period_Re12000_5}
\includegraphics[width=1.9in,height=1.9in]{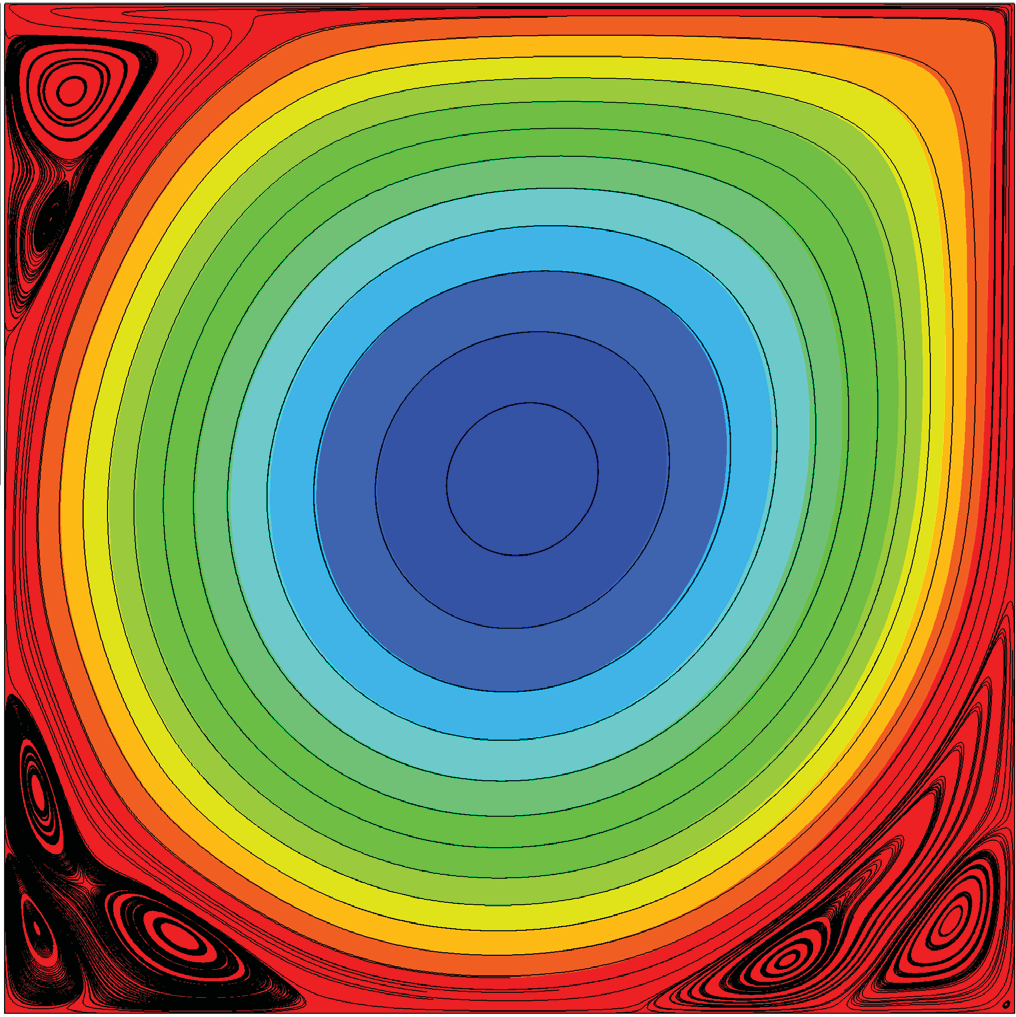}}
\subfigure[]{
\label{fig:LDF_Period_Re12000_6}
\includegraphics[width=1.9in,height=1.9in]{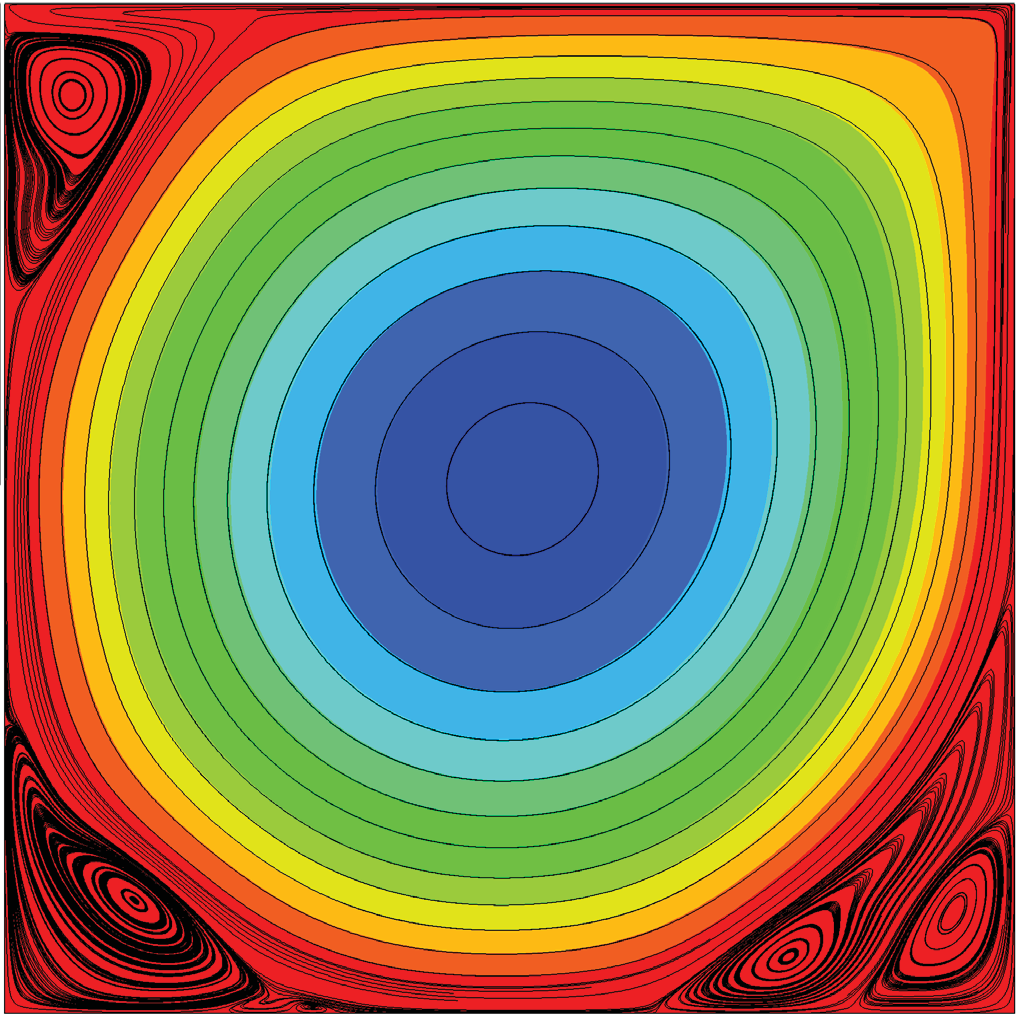}}
\caption{Stream-function contours and streamlines at $Re=12000$ in different times of a cycle: (a) $t=0$, (b) $t=T/5$, (c) $t=2T/5$, (d) $t=3T/5$, (e) $t=4T/5$, (f) $t=T$, } 
\label{fig:LDF_Period_Re12000}
\end{figure}

\begin{figure}
\subfigure[]{
\label{fig:LDF_hRe_Phase_46}
\includegraphics[width=2.7in,height=2.3in]{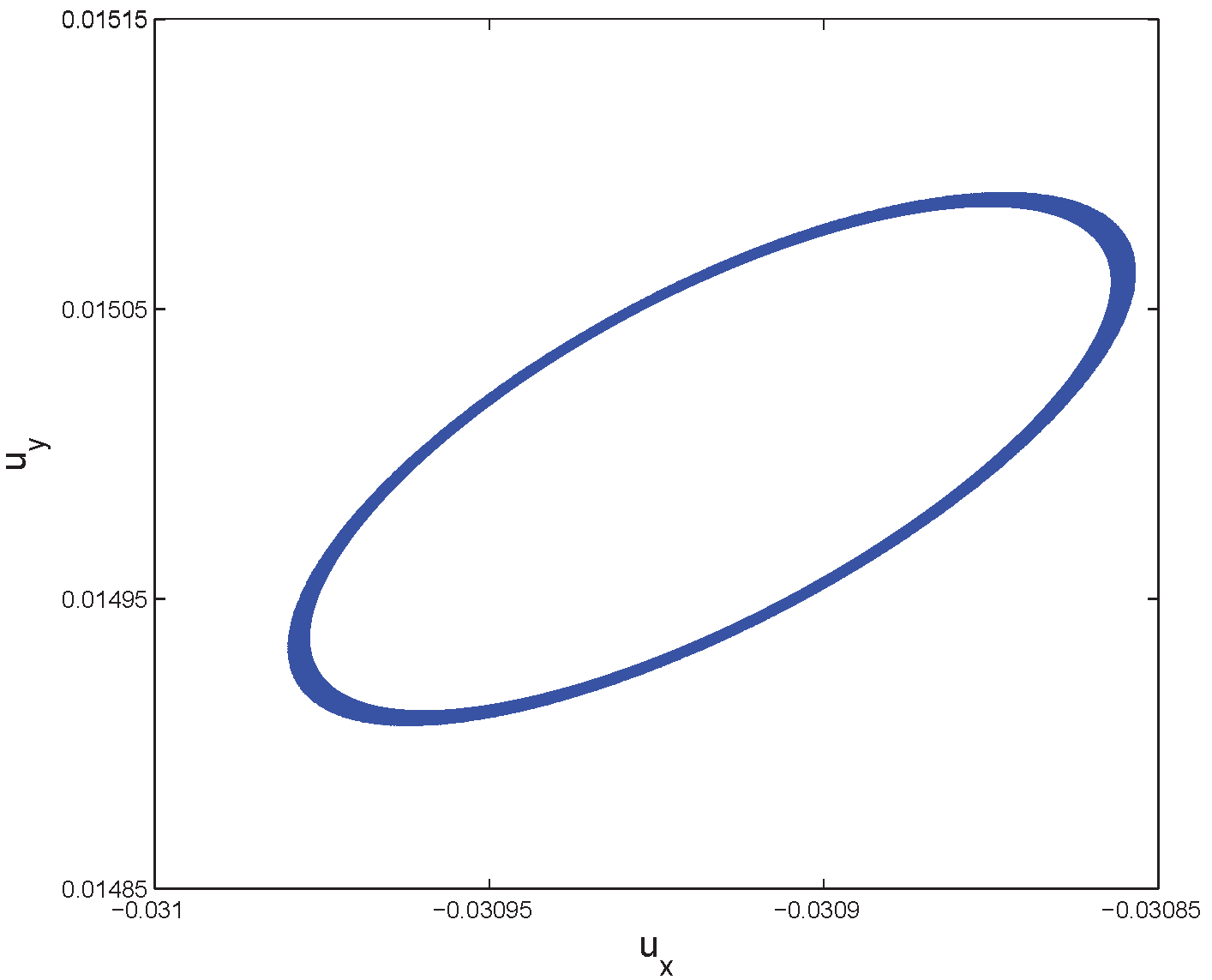}}
\subfigure[]{
\label{fig:LDF_hRe_Phase_87}
\includegraphics[width=2.7in,height=2.3in]{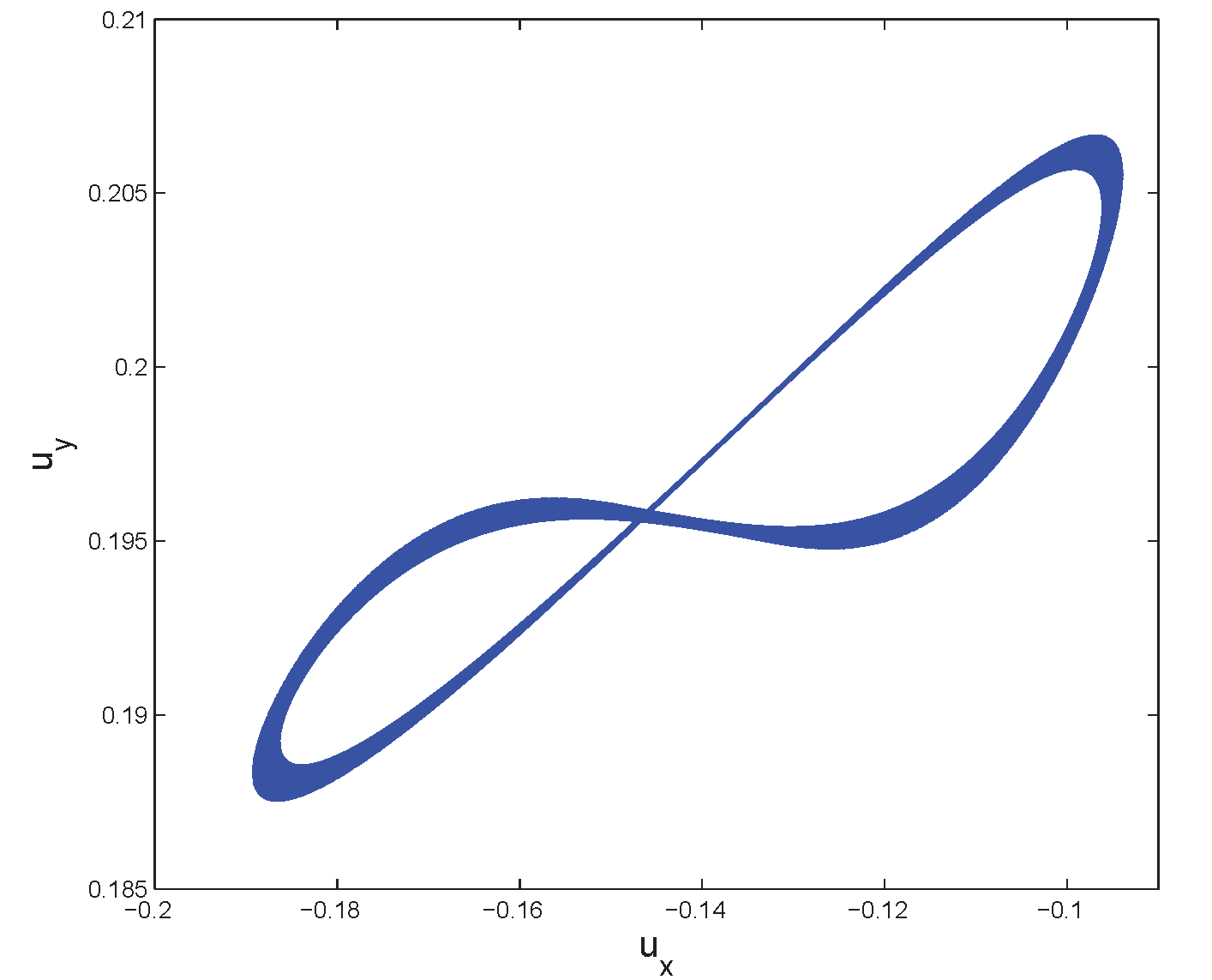}}
\caption{Phase-space trajectories of velocity at monitoring point $(0.494, 0.494)$ and $(0.220, 0.087)$ for $Re = 12000$.} 
\label{fig:LDF_hRe_Phase}
\end{figure}

As shown in Fig. \ref{fig:LDF_hRe}, the periodic solution with main frequency $f=0.6144$ is observed by monitoring the kinetic energy which is defined as ${E_k} = 0.5\int_\Omega  {{{\left\| u \right\|}^2}} dx$. And the creation, motion and merging of eddies near corners in a cycle is shown in Fig. \ref{fig:LDF_Period_Re12000}. The phase-space trajectories of velocity at monitoring points near center and near left corner are shown in Fig. \ref{fig:LDF_hRe_Phase}, and we noticed that the trajectories are not exact single lines which implies that the solution is not purely periodic. These results agree well with those reported by Cazemier et al. \cite{cazemier1998proper} and is similar to the result at $Re = 10000$ reported by Bruneau et al. \cite{bruneau20062d}. In addition, we also use the $N \times N = 80 \times 80$ and $N \times N = 96 \times 96$ nonuniform meshes for simulating this problem, but the computing results are steady rather than periodic. It implies that with the Reynold number increasing, the nonuniform mesh should be refined correspondingly as the uniform mesh does.

Consequently, the proposed model has good accuracy and robustness in simulating 2D-LDF problems by using proper nonuniform mesh. At high Reynold number ($Re=12000$), though the size of mesh is small, the proposed model can predict the complex phenomenon efficiently as shown above. Meanwhile, the refined nonuniform mesh is also needed to obtain the accurate result.

\subsection{3D lid-driven flow}
\label{sec:3D_LDF}
As the 2D lid-driven flow, 3D lid-driven flow (3D-LDF) is also a popular benchmark problem for testing numerical methods \cite{prasad1989reynolds, albensoeder2001three, leriche2000direct, albensoeder2005accurate, ku1987pseudospectral, feldman2010oscillatory, sheu2002flow}. For the proposed model can be extended to three dimensional flows conveniently, just like the original DUGKS model, we simulate the 3D-LDF at $Re = 400, 1000, 1900$ for verifying the incompressible DUGKS model.

\begin{figure}
\includegraphics[width=2.9in,height=2.7in]{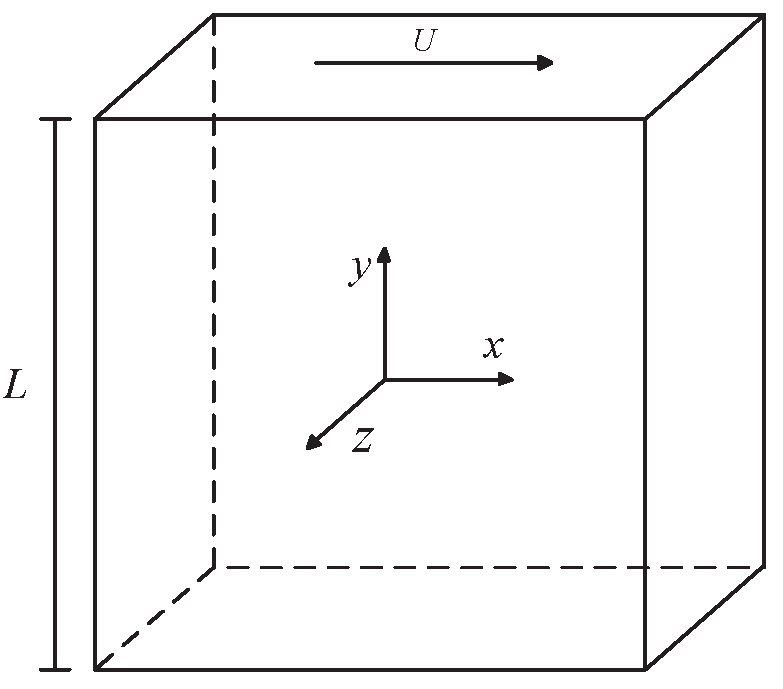}
\caption{Geometry of 3D lid-driven flow.} 
\label{fig:LDF_3D_sketch}
\end{figure}

The geometry of 3D-LDF is shown in Fig. \ref{fig:LDF_3D_sketch}. The length of cavity is $L=1.0$, and the driven velocity is set to be $U=1.0$ in $x$-direction. The parameters are set as follows: $RT = 100/3$ and the CFL number is set to be $0.5$. The equilibrium distribution is initialized by $u_x = 0$, $u_y = 0$, $u_z = 0$ and $\rho = 1$. Besides, the criterion of convergence in Sec. \ref{sec:Periodic_flow} is extended to three dimensional flows and employed to this problem.

\begin{figure}
\includegraphics[width=2.7in,height=2.2in]{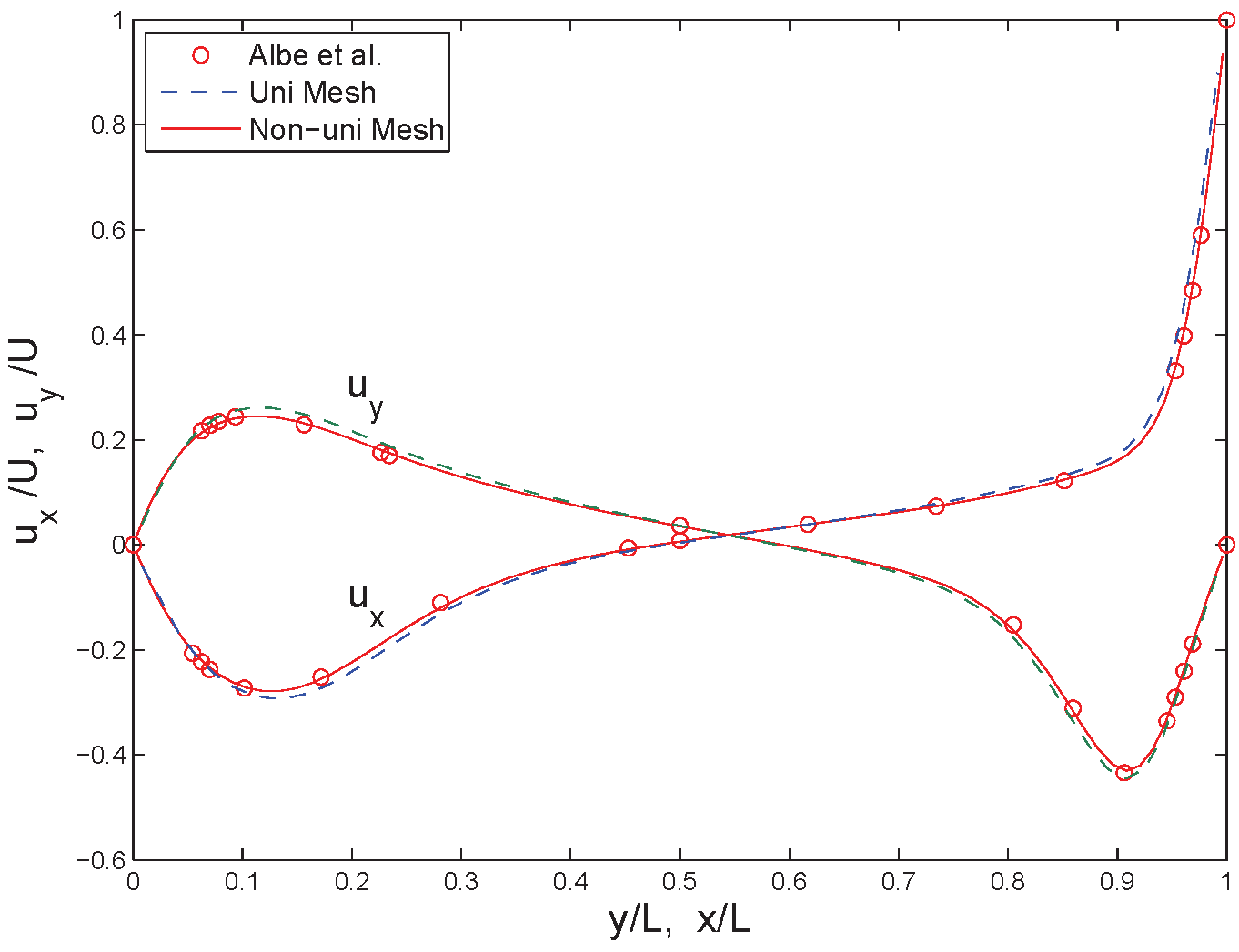}
\caption{Comparison of velocity profiles along the centerlines $(x, 0.5, 0.5)$ and $(0.5, y, 0.5)$ with reference data \cite{albensoeder2005accurate} at $Re = 1000$, where the solid lines are numerical results on nonuniform mesh, the dashed lines are numerical results on nonuniform mesh, and ($\circ$) is reference data.} 
\label{fig:3DLDF_U_N}
\end{figure}

At the beginning, we compared our results at $Re = 1000$ using the $60 \times 60 \times 60$ uniform and nonuniform meshes with the benchmarks of Albensoeder et al. \cite{albensoeder2005accurate} to validate the proposed model. By extending Eq. \eqref{Nonuniform_mesh} to three dimensional problems, the nonuniform mesh points can be generated. In Fig. \ref{fig:3DLDF_U_N}, the velocity profiles along the centerlines $(x, 0.5, 0.5)$ and $(0.5, y, 0.5)$ are shown, and our results have good agreement with the reference data. However, the data of nonuniform mesh is more close to the benchmark than that of uniform mesh when they are in same size. It is consistent with the conclusion in 2D-LDF which indicates that the nonuniform mesh has better performance in LDF problems. Thus, the nonuniform mesh will be used to study the following simulations.

\begin{figure}
\subfigure[]{
\label{fig:LDF_3DLDF_400}
\includegraphics[width=2.0in,height=1.9in]{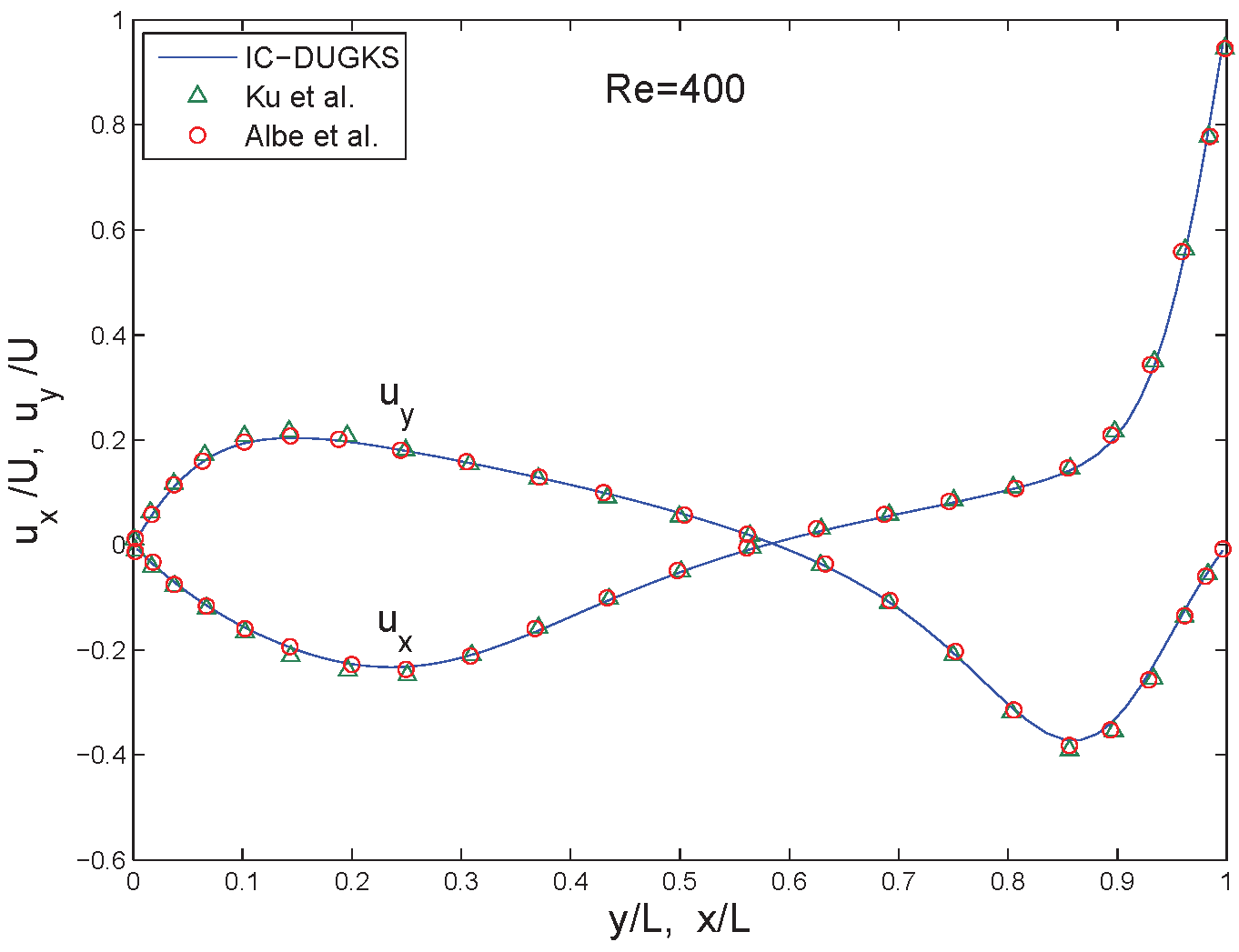}}
\subfigure[]{
\label{fig:LDF_3DLDF_1000}
\includegraphics[width=2.0in,height=1.9in]{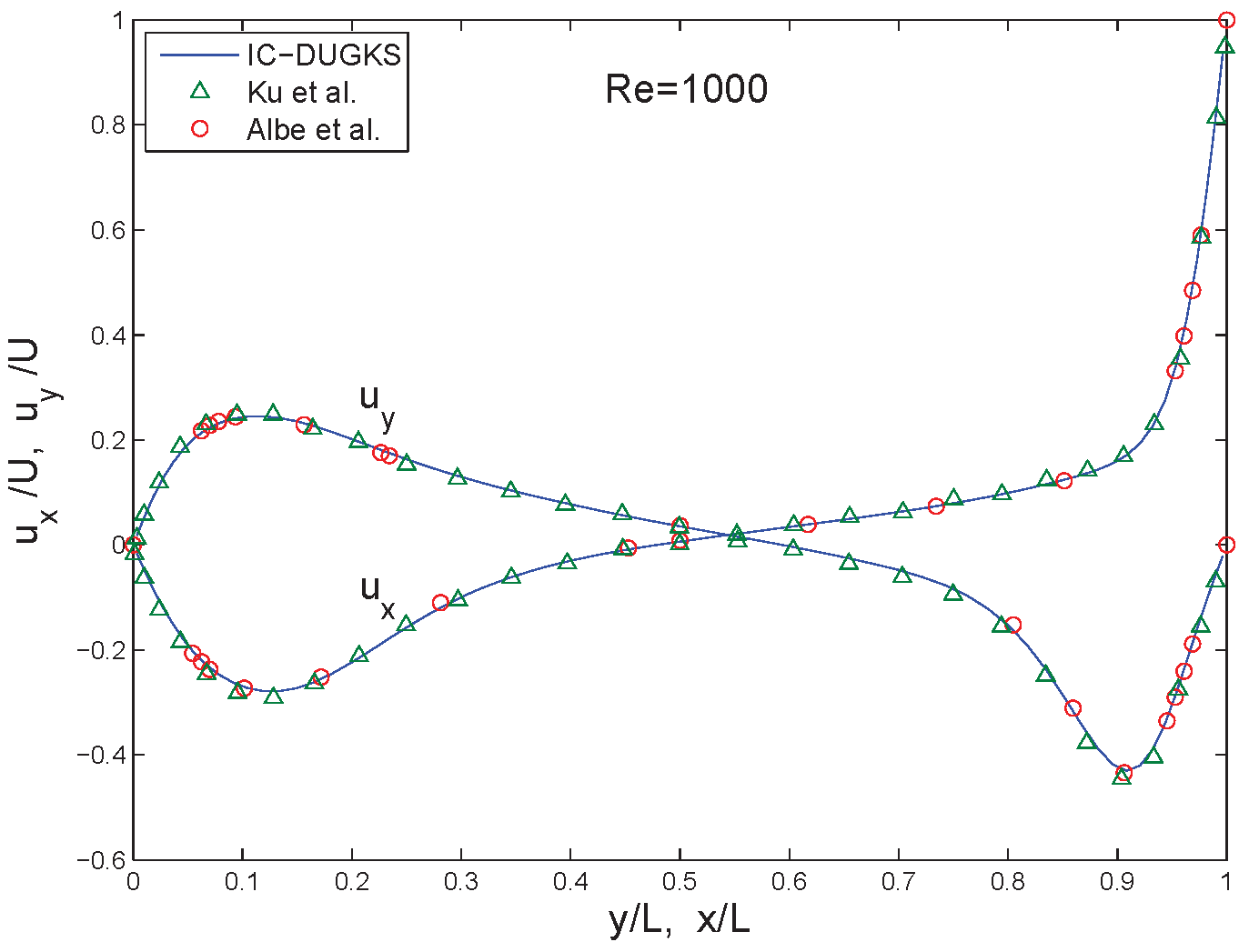}}
\subfigure[]{
\label{fig:LDF_3DLDF_1900}
\includegraphics[width=2.0in,height=1.9in]{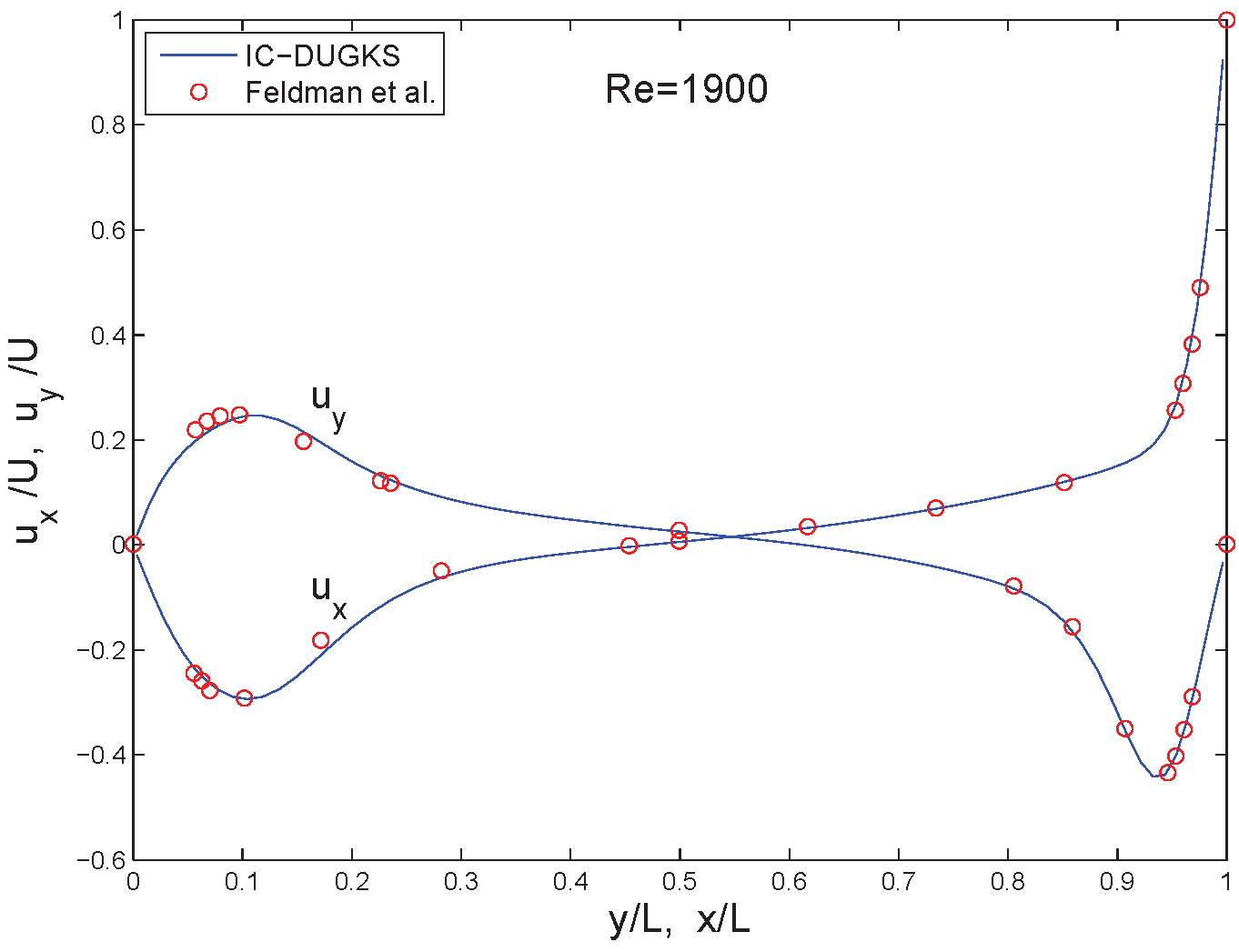}}
\caption{Comparison of velocity profiles along the centerlines $(x, 0.5, 0.5)$ and $(0.5, y, 0.5)$ with reference data \cite{ku1987pseudospectral, albensoeder2005accurate, feldman2010oscillatory}  at $Re = 400, 1000, 1900$ using $60 \times 60 \times 60$ nonuniform mesh. (\cite{ku1987pseudospectral, feldman2010oscillatory} are extracted from their figures.)} 
\label{fig:3D_LDF_3Re}
\end{figure}

Furthermore, Fig. \ref{fig:3D_LDF_3Re} shows the comparison of the benchmarks \cite{ku1987pseudospectral, albensoeder2005accurate, feldman2010oscillatory} and our data using the $60 \times 60 \times 60$ nonuniform mesh at $Re = 400, 1000, 1900$. With the Reynold number increasing, the numerical results maintain good agreement with the reference data. Especially, at $Re = 1900$, the benchmark \cite{feldman2010oscillatory} is computed by $152 \times 152 \times 152$ and $200 \times 200 \times 200$ meshes, yet the size of our mesh is far less. This fact suggests that the proposed model are accurate and efficient in simulating 3D-LDF problem.

\begin{figure}
\includegraphics[width=2.7in,height=2.4in]{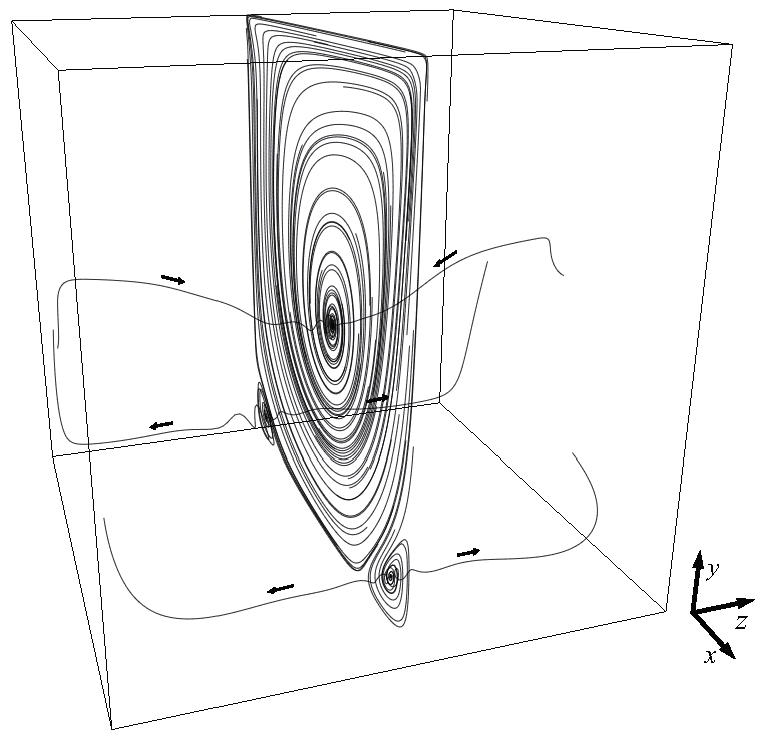}
\caption{Streamlines near the main and secondary eddies at $z = 0.5$. The lid velocity is along $x$-direction.} 
\label{fig:LDF_3D_streamline_benchmark_m}
\end{figure}

The flow topology of 3D-LDF is remarkable. As the result shows above, when the Reynold number at $Re = 1000$, the flow will reach the steady state. In Fig. \ref{fig:LDF_3D_streamline_benchmark_m}, the streamlines in the cavity midplane $z = 0.5$ never left the plane, and streamlines started from the points which near the centers of the main and secondary eddies stretch strongly in $z$-direction with symmetries about the midplane. This phenomenon is also reported by Feldman et al. \cite{feldman2010oscillatory}. 

\begin{figure}
\subfigure[]{
\label{fig:LDF_3D_limiting_streamline_with_p}
\includegraphics[width=2.7in,height=2.3in]{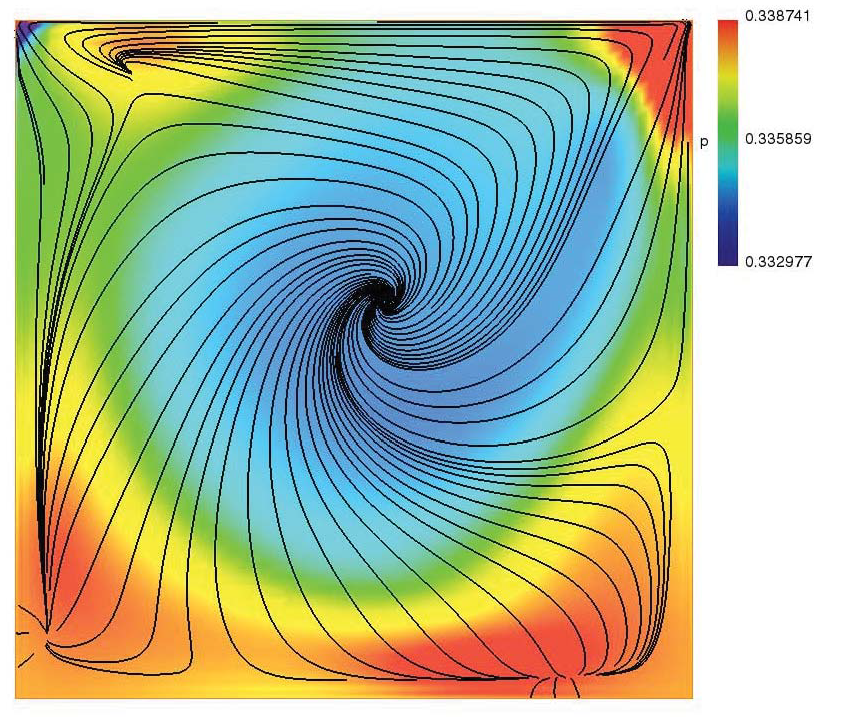}}
\subfigure[]{
\label{fig:LDF_3D_limiting_streamline_with_mag}
\includegraphics[width=2.7in,height=2.3in]{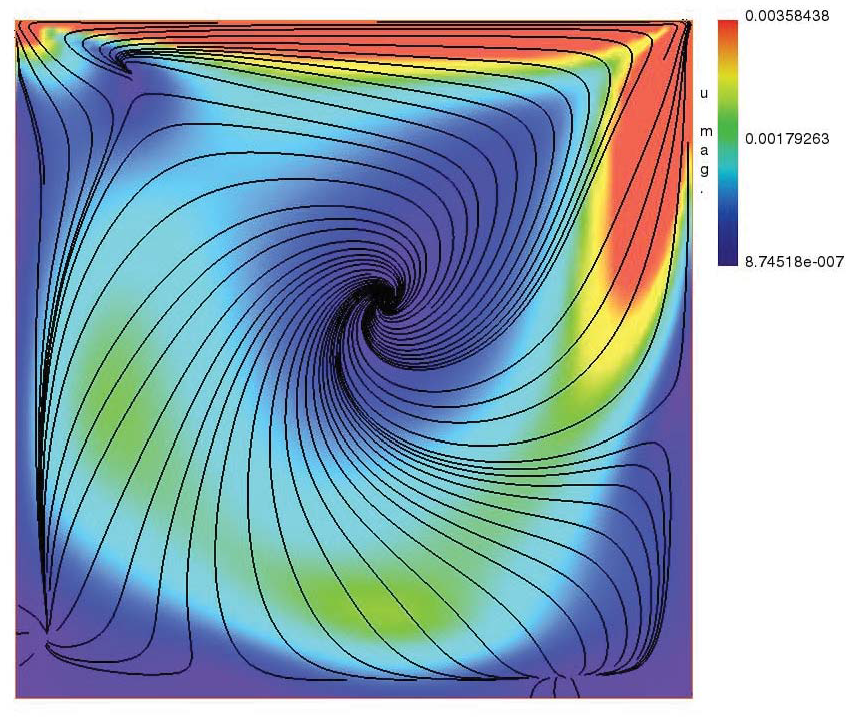}}
\caption{Limiting streamlines and contours at $z=1$. (a) is the pressure contour, and (b) is the velocity magnitude contour.} 
\label{fig:LDF_3D_limiting_streamline}
\end{figure}

\begin{figure}
\includegraphics[width=2.9in,height=2.4in]{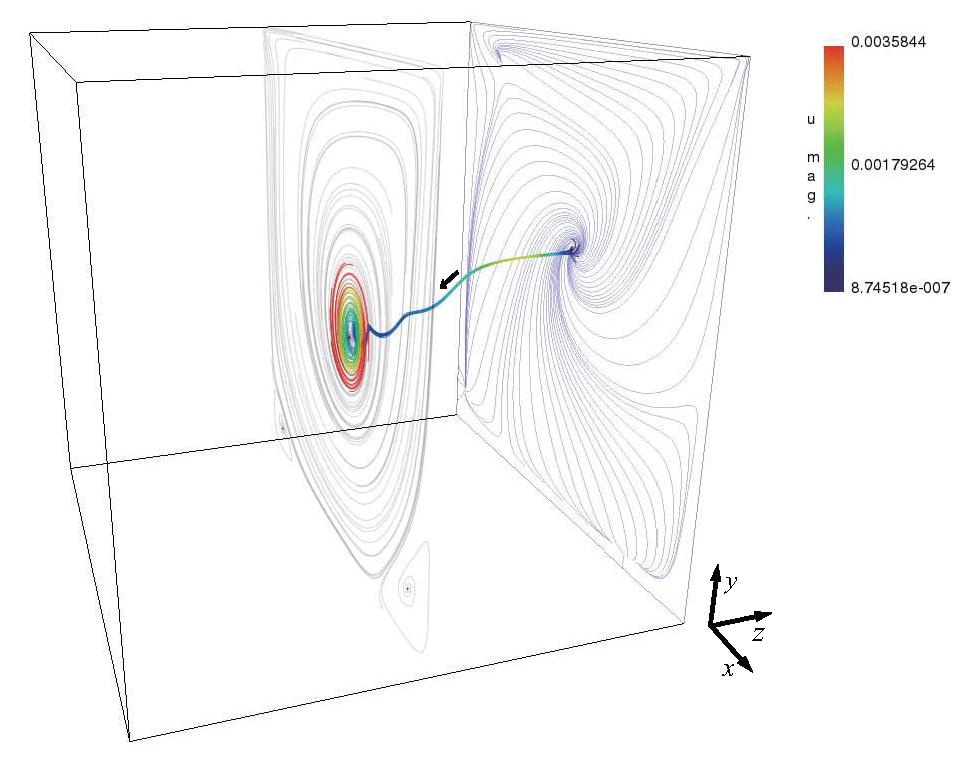}
\caption{Streamlines near the focus point and the limiting streamlines at $z=1$} 
\label{fig:LDF_3D_streamline_wall_m}
\end{figure}

With the limiting streamlines (L-streamlines) which is defined as streamlines immediately above the surface \cite{chiang1997numerical}, the topological studies at the cavity side walls $z = 1$ ($z = 0$ has the symmetric results) are presented. Fig. \ref{fig:LDF_3D_limiting_streamline} shows the contours of pressure and velocity magnitude with the L-streamlines. Around the geometry center, the area of lower pressure and velocity is found, and a focus point $(0.537, 0.583, 1)$, where the L-streamlines spiral into this node \cite{sheu2002flow}, is also shown in the area. That implies when the fluid particles located in this area, they will be attracted to the focus point. Back to the streamlines, as shown in Fig. \ref{fig:LDF_3D_streamline_wall_m}, they imply that the fluid particles near the focus point directly move to the midplane. At the same time, for the color of the streamlines represented the velocity magnitude, it should be noticed that the particles accelerate slightly (turns green) at the beginning, then decelerate (turns blue) near the mid plane, and finally, accelerate significantly (turns red) close to the midplane with spiraling motion. Therefore, combined the results of the L-streamlines and streamlines, the motion of the particles located around the focus point can be characterized. As the steady state of 3D-LDF is a sequence of the symmetric problem geometry \cite{feldman2010oscillatory}, we can predict the type of motion. And this phenomenon is also found at $Re = 400, 1900$, consequently. 

The position of the focus points at $Re = 400, 1000, 1900$ are given as $(0.623, 0.696, 1)$, $(0.537, 0.583, 1)$ and $(0.505, 0.585, 1)$, and as a reference, the result at $Re = 400$ computed by Sheu et al. \cite{sheu2002flow} is $(0.669, 0.746, 1)$ (extracted from their Fig. 7). 

\section{Conclusions}
In this paper, the incompressible DUGKS model with external force term is proposed for incompressible fluid flows. The advantages of the original DUGKS model is preserved, such as the simplified flux evaluation scheme, conservative collision operator and the asymptotic preserving properties. Besides, we also introduced the NEQ scheme for both velocity and pressure boundary conditions into DUGKS. The proposed model has been verified by several test cases, and the numerical solutions achieve excellent agreements with both analytical and benchmark results. Through the validations, it can be found that the proposed model is of second-order accuracy and can reduce the compressible errors significantly. Also, the NEQ scheme performs very well as in the same convergence order with the proposed model, and the application of non-uniform mesh for improving the computational efficiency is successful. Additionally, the incompressible DUGKS model is implemented in simulating three dimensional flows. Compared with the benchmarks, the numerical solutions of cubical lid-driven flows show that the proposed model is also accurate and robust. And by analyzing the numerical solutions topologically, the motion pattern of the fluid particles near the focus point is found for the steady-state cubical lid-driven flows.

\appendix
\renewcommand{\thefigure}{A-\arabic{figure}}
\setcounter{figure}{0}
\section{APPENDIX: Algorithm of incompressible DUGKS with external force term}
\label{sec:appendix}

The algorithm of incompressible DUGKS with external force term is shown to illustrate the evolution of distribution function. The algorithm structure is same with the finite volume scheme, but the evolution is of the distribution function rather than the macro values such as density and velocity. For instances, the evolution algorithm is given with the rectangular mesh, and it can be extended to unstructured mesh just as the general finite volume scheme.

\begin{figure}
\includegraphics[width=3.2in,height=2.4in]{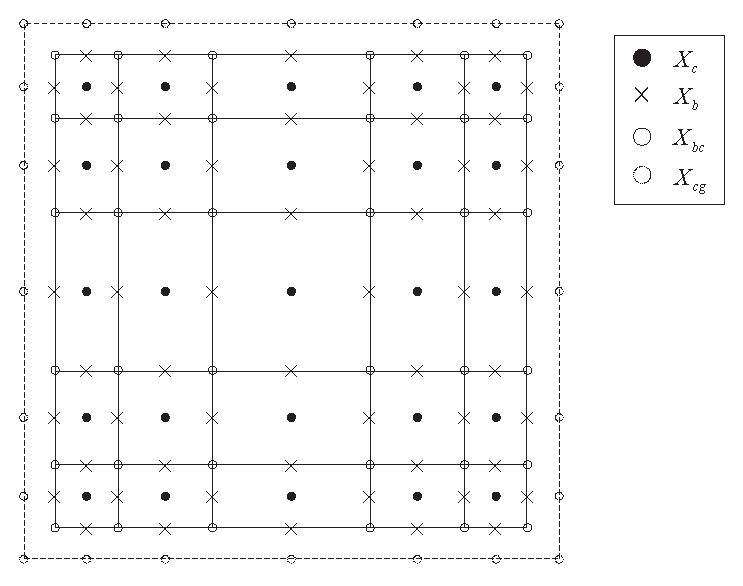}
\caption{Geometry of the rectangular mesh.} 
\label{fig:mesh}
\end{figure}

The geometry of the mesh is shown in Fig. \ref{fig:mesh}, where $\bm{x}_i \in X_c$ is the control volume center point, $\bm{x}_i \in X_b$ is the point located in the middle of the interface, $\bm{x}_i \in X_{bc}$ is the crosspoint between the interfaces, and $\bm{x}_i \in X_{cg}$ is the ghost point \cite{leveque2002finite} which is used to simplify the interpolation at boundary. Thus, the set of all the mesh points is $X = X_c \cup X_b \cup X_{bc} \cup X_{cg}$. The $f(\bm{x}_i, \bm{\xi}_k)$ and  $F(\bm{x}_i, \bm{\xi}_k)$ are the memory space saving the distribution function at point $\bm{x}_i$ with the discrete particle velocity $\bm{\xi}_k$. Besides, the density $\rho_i$ and the velocity $\bm{u}_i$ are short for $\rho(\bm{x}_i)$ and $\bm{u}(\bm{x}_i)$. The algorithm below illustrates that how to update the distribution function $F(\bm{x}_i, \bm{\xi}_k)$ from time $t$ to $t + \Delta t$.

\begin{algorithm}[H]
\renewcommand{\thealgorithm}{} 
\caption{Evolution of the distribution function}
\label{alg:aco}
\begin{algorithmic}[0]
\State \\ \# 1. Compute $\bar f^{+}(\bm{x}_i, \bm{\xi}_k, t)$ at the cell center points as Eq. \eqref{f_bar_+}.
\ForAll {$\bm{x}_i \in X_c$ and $\bm{\xi}_k$}
\State $f(\bm{x}_i, \bm{\xi}_k) \leftarrow \frac{{2\tau  - h}}{{2\tau  + \Delta t}}F(\bm{x}_i, \bm{\xi}_k)+ \frac{{3h}}{{2\tau  + \Delta t}}{f^{eq}(\bm{\xi}_k, \rho_i, \bm{u}_i)} + \frac{{3\tau h}}{{2\tau  + \Delta t}}S(\bm{\xi}_k, \rho_i, \bm{u}_i),$
\State \# where the $f^{eq}(\bm{\xi}_k, \rho_i, \bm{u}_i)$ and $S(\bm{\xi}_k, \rho_i, \bm{u}_i)$ are computed from Eq. \eqref{icfeq} and Eq. \eqref{S}.
\EndFor

\State \# 2. Compute $\bar f^{+}(\bm{x}_i, \bm{\xi}_k, t)$ at the ghost points by extrapolation
\ForAll {$\bm{x}_i \in X_{cg}$ and $\bm{\xi}_k$}
\State $f(\bm{x}_i, \bm{\xi}_k) \leftarrow$ extrapolation from $f(\bm{x}_j, \bm{\xi}_k)$,
\State \# where $\bm{x}_j \in X_c$ are several points close to $\bm{x}_i$ depending on the boundary.
\EndFor

\State \# 3. Compute $\bar f^{+}(\bm{x}_i, \bm{\xi}_k, t)$ at the interface points by center interpolation
\ForAll {$\bm{x}_i \in X_{b} \cup X_{bc}$ and $\bm{\xi}_k$}
\State $f(\bm{x}_i, \bm{\xi}_k) \leftarrow$ center interpolation from $f(\bm{x}_j, \bm{\xi}_k)$,
\State \# where $\bm{x}_j \in X_c \cup X_{cg}$ are several points around $\bm{x}_i$.
\EndFor

\algstore{DUGKS} 
\end{algorithmic} 
\end{algorithm} 

\begin{algorithm}[H]
\begin{algorithmic}[0]
\algrestore{DUGKS} 

\State \\ \# 4. Compute $\bar f(\bm{x}_i, \bm{\xi}_k, t + h)$ at the interface points as Eq. \eqref{gradient}
\ForAll {$\bm{x}_i \in X_{b}$ and $\bm{\xi}_k$}
\State $\bm{\sigma _b} \leftarrow$ difference of $f(\bm{x}_j, \bm{\xi}_k)$,
\State \# where $\bm{x}_j \in X_c \cup X_{bv} \cup X_{cg}$ are several points close to $\bm{x}_i$ depending on gradient direction.
\State $F(\bm{x}_i, \bm{\xi}_k) \leftarrow f(\bm{x}_i, \bm{\xi}_k) - \bm{\xi}_k h \cdot \bm{\sigma _b}$,
\EndFor

\State \# 5. Boundary processing of $\bar f(\bm{x}_i, \bm{\xi}_k, t + h)$ ($\bm{x}_i$ located at cell interface)
\ForAll {$\bm{x}_i$ at boundaries and $\bm{\xi}_k$ required by boundary condition}
\State $F(\bm{x}_i, \bm{\xi}_k) \leftarrow F(\bm{x}_i, \bm{\xi}_k)$ processed by boundary condition, 
\State \# such as BB with Eq. \eqref{BB} or NEQ with Eq. \eqref{NEQ}.
\EndFor

\State \# 6. Compute $\rho(\bm{x}_i)$ and $\bm{u}(\bm{x}_i)$ at $x_b$ as Eq. \eqref{macro_values}
\ForAll {$\bm{x}_i \in X_{b}$ and $\bm{\xi}_k$}
\State $\rho(\bm{x}_i) \leftarrow \sum\limits_k {F(\bm{x}_i, \bm{\xi}_k)}$

\State $\bm{u}(\bm{x}_i) \leftarrow \frac{1}{\rho_0} \left( \sum\limits_k {\bm{\xi}_kF(\bm{x}_i, \bm{\xi}_k)} + \frac{{\rho_0 Gh}}{2}\right)$
\EndFor

\State \# 7. Compute the original distribution function  $f(\bm{x}_i, \bm{\xi}_k, t + h)$ at interface as Eq. \eqref{original_f}
\ForAll {$\bm{x}_i \in X_{b}$ and $\bm{\xi}_k$}
\State $F(\bm{x}_i, \bm{\xi}_k) \leftarrow \frac{{2\tau }}{{2\tau  + h}}F(\bm{x}_i, \bm{\xi}_k) + \frac{h}{{2\tau  + h}}{f^{eq}(\bm{\xi}_k, \rho_i, \bm{u}_i)} + \frac{{\tau h}}{{2\tau  + h}}S(\bm{\xi}_k, \rho_i, \bm{u}_i)$
\EndFor

\State \# 8. Compute the micro flux from $f(\bm{x}_i, \bm{\xi}_k, t + h)$ as Eq. \eqref{microflux}
\ForAll {$\bm{x}_i \in X_{c}$ and $\bm{\xi}_k$}
\State ${m(\bm{x}_i)} \leftarrow \sum_{\partial {V_i}} {(\bm{\xi}_k  \cdot \bm{n}_j)F(\bm{x}_j, \bm{\xi}_k))} {\kern 1pt} s_j,$
\State \# where $\bm{x}_j \in X_b$ are the interface points around the cell center $\bm{x}_i$,
\State \# $\bm{n}_j$ and $s_j$ are the outward unit vector and the area (length) of the corresponding interface.
\EndFor

\algstore{DUGKS1} 
\end{algorithmic} 
\end{algorithm} 

\begin{algorithm}[H]
\begin{algorithmic}[0]
\algrestore{DUGKS1} 

\State \# 9. Update $\tilde f(\bm{x}_i, \bm{\xi}_k, t + \Delta t)$ at cell center as Eq. \eqref{evolution_equation}
\ForAll {$\bm{x}_i \in X_{c}$ and $\bm{\xi}_k$}
\State $\tilde f^+(\bm{x}_i, \bm{\xi}_k) \leftarrow \frac{{4}}{{3}}f(\bm{x}_i, \bm{\xi}_k) - \frac{{1}}{{3}} F(\bm{x}_i, \bm{\xi}_k)$
\State \# compute $\tilde f^+$ as Eq. \eqref{f_tilde_+}
\State $F(\bm{x}_i, \bm{\xi}_k) \leftarrow \tilde f^+(\bm{x}_i, \bm{\xi}_k) - \frac{{\Delta t}}{{\left| {{V_i}} \right|}}{m(\bm{x}_i)}$
\EndFor

\State \# 10. Update $\rho(\bm{x}_i)$ and $\bm{u}(\bm{x}_i)$ at cell center as Eq. \eqref{macro_values}
\ForAll {$\bm{x}_i \in X_{c}$ and $\bm{\xi}_k$}
\State $\rho(\bm{x}_i) \leftarrow \sum\limits_k {F(\bm{x}_i, \bm{\xi}_k)}$

\State $\bm{u}(\bm{x}_i) \leftarrow \frac{1}{\rho_0} \left( \sum\limits_k {\bm{\xi}_kF(\bm{x}_i, \bm{\xi}_k)} + \frac{{\rho_0 G\Delta t}}{2}\right)$
\EndFor

\end{algorithmic} 
\end{algorithm} 

As we can find that the distribution function $\tilde f(\bm{x}_i, \bm{\xi}_k, t)$ at cell center saved in $F(\bm{x}_i, \bm{\xi}_k)$ is updated from time $t$ to $t + \Delta t$, and the density $\rho(\bm{x}_i)$ and velocity $\bm{u}(\bm{x}_i)$ are also updated. To obtain the results, we just need to repeat this loop and end it with proper conditions.

\bibliographystyle{unsrt}
\bibliography{Bibliography}

\end{document}